\documentclass[11pt,a4paper]{article}

\usepackage{amsmath,amsfonts,amssymb,amsthm,amstext,amscd,array}
\usepackage{tikz}
\usepackage{mathrsfs}
\usepackage{bbm}
\usepackage{bm}
\usepackage[utf8]{inputenc}
\usepackage[arrow,matrix,curve]{xy}

\newcommand{\C}{\complex}
\newcommand{\Z}{\zed}

\newcommand{\mbf}[1]{{\boldsymbol {#1} }}
\def\ii{{\,{\rm i}\,}}
\def\dd{{\rm d}}
\def\hh{{\rm h}}

\def\a{{\sf a}}

\def\sff{{\sf f}}

\newcommand{\unit}{\mathbbm{1}}   			

\newcommand{\Hom}{{\rm Hom}}

\newcommand{\frg}{\mathfrak{g}}				
\newcommand{\frh}{\mathfrak{h}}				

\newcommand{\CCM}{\mathscr{M}}
\newcommand{\CCR}{\mathscr{R}}

\newcommand{\CG}{\mathcal{G}}
\newcommand{\CH}{\mathcal{H}}
\newcommand{\CU}{\mathcal{U}}
\newcommand{\CM}{\mathcal{M}}
\newcommand{\Wcal}{\mathcal{W}}

\newcommand{\eq}{\begin{equation}}
\newcommand{\eqend}{\end{equation}}
\newcommand{\eqa}{\begin{eqnarray}}
\newcommand{\nonueqa}{\begin{eqnarray*}}
\newcommand{\eqaend}{\end{eqnarray}}
\newcommand{\nonueqaend}{\end{eqnarray*}}

\newcommand{\bma}[1]{\begin{array}{#1}}
\newcommand{\ema}{\end{array}}
\newcommand{\bc}{\begin{center}}
\newcommand{\ec}{\end{center}}

\newcommand{\R}{\real}

\renewcommand{\thefootnote}{\fnsymbol{footnote}}
\newcommand{\newsection}{\setcounter{equation}{0} \section}

\newcommand{\complex}{{\mathbb C}} 
\newcommand{\zed}{{\mathbb Z}} 
\newcommand{\real}{{\mathbb R}} 

\def\alg{{\mathcal A}}

\newif\ifold             \oldtrue

\def\e{{\,\rm e}\,}

\def\be{\begin{equation}}
\def\ee{\end{equation}}
\def\bea{\begin{eqnarray}}
\def\eea{\end{eqnarray}}
\def\bd{\begin{displaymath}}
\def\ed{\end{displaymath}}

\newcommand{\beq}{\begin{eqnarray}}
\newcommand{\eeq}{\end{eqnarray}}

\makeatletter
\newdimen\normalarrayskip              
\newdimen\minarrayskip                 
\normalarrayskip\baselineskip
\minarrayskip\jot
\newif\ifold             \oldtrue            
\def\arraymode{\ifold\relax\else\displaystyle\fi} 
\def\@arrayskip{\ifold\baselineskip\z@\lineskip\z@
     \else
     \baselineskip\minarrayskip\lineskip2\minarrayskip\fi}
\def\@arrayclassz{\ifcase \@lastchclass \@acolampacol \or
\@ampacol \or \or \or \@addamp \or
   \@acolampacol \or \@firstampfalse \@acol \fi
\edef\@preamble{\@preamble
  \ifcase \@chnum
     \hfil$\relax\arraymode\@sharp$\hfil
     \or $\relax\arraymode\@sharp$\hfil
     \or \hfil$\relax\arraymode\@sharp$\fi}}
\def\@array[#1]#2{\setbox\@arstrutbox=\hbox{\vrule
     height\arraystretch \ht\strutbox
     depth\arraystretch \dp\strutbox
     width\z@}\@mkpream{#2}\edef\@preamble{\halign \noexpand\@halignto
\bgroup \tabskip\z@ \@arstrut \@preamble \tabskip\z@ \cr}%
\let\@startpbox\@@startpbox \let\@endpbox\@@endpbox
  \if #1t\vtop \else \if#1b\vbox \else \vcenter \fi\fi
  \bgroup \let\par\relax
  \let\@sharp##\let\protect\relax
  \@arrayskip\@preamble}
\makeatother

\usepackage{hyperref}
\hypersetup{
pdftitle={Equivariant Dimensional Reduction},pdfauthor={Richard Szabo and Omar Valdivia}}

\begin{document}
\begin{titlepage}
\begin{flushright}

\baselineskip=12pt

EMPG--14--7
\end{flushright}

\begin{center}

\vspace{1cm}

\baselineskip=24pt

{\Large\bf Covariant Quiver Gauge Theories}

\baselineskip=14pt

\vspace{1cm}

{\bf Richard
  J. Szabo}${}^{1,}$\footnote{Email: \ {\tt R.J.Szabo@hw.ac.uk}} \ and \ {\bf Omar Valdivia}${}^{1,2,}$\footnote{Email: \ {\tt ov32@hw.ac.uk}}
\\[4mm]
\noindent ${}^1$ {\it Department of Mathematics\\ Heriot--Watt University\\ Colin Maclaurin Building,
  Riccarton, Edinburgh EH14 4AS, U.K.}\\
\noindent and {\it Maxwell Institute for
Mathematical Sciences, Edinburgh, U.K.}\\
\noindent and {\it The Tait Institute, Edinburgh, U.K.}
\\[4mm]
\noindent ${}^2$ {\it Departamento de F\'isica\\ Universidad de Concepci\'on\\ Casilla 160-C,
  Concepci\'on, Chile}
 \\[30mm]

\end{center}

\begin{abstract}

\baselineskip=12pt

We consider dimensional reduction of gauge theories with arbitrary
gauge group in a formalism based on equivariant principal bundles. For
the classical gauge groups we clarify the relations between
equivariant principal bundles and quiver bundles, and show that the
reduced quiver gauge theories are all generically built on the same
universal symmetry breaking pattern. The formalism enables the
dimensional reduction of Chern--Simons gauge theories in arbitrary odd
dimensionalities. The reduced model is a novel Chern--Simons--Higgs
theory consisting of a Chern--Simons term valued in the residual gauge
group plus a higher order gauge and diffeomorphism invariant coupling of
Higgs fields with the gauge fields. We study the moduli spaces of solutions, which in some instances provide geometric representations of certain quiver varieties as moduli spaces of flat invariant connections. As physical applications, we
consider dimensional reductions involving non-compact gauge supergroups as
a means for systematically inducing novel couplings between gravity
and matter. In particular, we reduce Chern--Simons
supergravity to a quiver gauge theory of AdS gravity
involving a non-minimal coupling to scalar Higgs fermion fields.

\end{abstract}

\end{titlepage}
\setcounter{page}{2}

\newpage

{\baselineskip=12pt
\tableofcontents
}

\allowdisplaybreaks

\bigskip

\renewcommand{\thefootnote}{\arabic{footnote}}
\setcounter{footnote}{0}

\newsection{Introduction and summary}

Dimensional reduction provides a means of unifying gauge and Higgs sectors into a pure Yang--Mills theory in higher dimensions. The reductions are particularly rich if the extra spacetime
dimensions admit isometries, which can then be implemented on gauge orbits of fields~\cite{ForgaCS:1979zs}. The natural setting for spacetime isometries are coset
spaces $G/H$ of compact Lie groups in which Yang--Mills theory on the product space $ M\times
G/H$ is reduced to a Yang--Mills--Higgs theory on the manifold $M$; the construction can be 
extended supersymmetrically and also embedded in string theory \cite{Kapetanakis:1992hf}.
Equivariant dimensional reduction is an alternative approach
which naturally incorporates background fluxes coming from the topology of the canonical connections on the principal $H$-bundle $G\to G/H$~\cite{AlGar12,Lechtenfeld:2007st,Dolan:2010ur}; the reduced Yang--Mills--Higgs model is then succinctly described by a quiver gauge theory on $M$ whose underlying quiver is canonically associated to the representation theory of the Lie groups $H\subset G$. Such reductions have been used to
describe vortices as generalized instantons in higher-dimensional
Yang--Mills theory \cite{AlvarezConsul:2001mb,Lechtenfeld:2006wu,Popov:2007ms,Lechtenfeld:2008nh,Popov:2010rf}, as well as to construct explicit
$SU(2)$-equivariant monopole and dyon solutions of pure Yang--Mills theory in
four dimensions~\cite{Popov:2008wh}. 

A related approach is described in \cite{Manton:2010mj} which systematically translates the inverse relations of restriction and induction of vector bundles \cite{AlGar12} into the framework of
principal bundles. In this formulation there is no restriction on the
structure group and it permits, for instance, the application of equivariant dimensional
reduction techniques to gauge theories involving arbitrary gauge groups
$\mathcal{G}$. In the following we adapt such an approach to the simplest case where
$G=SU(2)  $ and $H=U(1)$, so that the internal coset
space $G/H$ is the two-sphere $S^2$ or the complex projective line $%
\mathbb{C}
P^{1}$. This example turns out to be rich enough to capture many of the general
features that one would encounter on generic cosets $G/H$.

The geometric structures arising from reductions of $SU(2)$-invariant Yang--Mills theory have been thoroughly studied in a multitude of
different contexts \cite{Popov:2008gw,Popov:2005ik,Dolan:2009ie}, while coset space dimensional reduction of five-dimensional Chern--Simons theory with gauge group $\CG=SU(2)$ is considered in~\cite{TempleRaston:1994sk} and some physical characteristics of its moduli space of solutions are pointed out. In
this paper we pursue the equivariant dimensional
reduction of topological gauge theories. We study the related problems of generalizing equivariant dimensional reduction to arbitrary
gauge groups $\mathcal{G}$ and extending these techniques to Chern--Simons gauge theories.
We classify the
symmetry breaking patterns induced by $G$-invariant connections whose gauge group $\CG$ lies in one of the four infinite families of classical Lie groups. We show that the unbroken gauge group of the reduced theory is generically the same (without any conditions on the
background Dirac monopole charges) in all cases. As a
consequence, the induced quiver gauge theories are the same for any classical gauge group (up to redefinitions of the coupling
constants); in our approach many of the geometric ingredients used
in~\cite{AlGar12,Lechtenfeld:2007st,Dolan:2010ur} to derive these
quiver gauge theories are translated into an algebraic framework. We will then explore $G$-equivariant dimensional
reduction of pure topological gauge theories. We calculate the reduction of an arbitrary odd-dimensional Chern--Simons
form over $%
\mathbb{C}
P^{1}$; although Chern--Simons Lagrangians are not gauge-invariant, we circumvent this problem by regarding them in the framework of transgression forms.
The reduced theory is a novel diffeomorphism-invariant
Chern--Simons--Higgs model, which can have local degrees of freedom whose dynamics and canonical structure are rather delicate to disentangle; our generally covariant models are therefore generically \emph{not} topological field theories.

As mathematical applications, we study the
moduli spaces of classical solutions of these field theories and
obtain some geometric interpretations of representation
theoretic results. For example, we describe the quiver varieties
parameterizing semisimple
representations of certain deformed preprojective algebras as moduli spaces of $SU(2)$-invariant flat
$\CG$-connections on the three-manifold $\CM=\R\times S^2$. As physical applications, we consider the case of non-compact gauge supergroups. In particular, we perform the dimensional reduction of five-dimensional
Chern--Simons supergravity over $\mathbb{C}P^1$. We show that if the
Higgs fields are bifundamental fields in the fermionic sector of the gauge algebra, then the reduced action contains the standard
Einstein--Hilbert term plus a non-minimal coupling of the Higgs fermions to the curvature.
This reduction scheme thus constitutes a novel systematic way to couple scalar fermionic fields to
gravitational Lagrangians, in a manner whereby non-vacuum solutions of three-dimensional AdS gravity can be lifted to give new solutions of five-dimensional supergravity on product spacetimes $M\times S^2$. 

This paper is organized as follows. In section \ref{sectionn2} we discuss general
aspects of $SU(2)$-equivariant dimensional reduction and
revisit the example of pure Yang--Mills theory as illustration. In section \ref{sector3} the
symmetry breaking patterns are analysed for the classical gauge groups
and the geometric structure of general principal quiver bundles is described.
 In section~\ref{sect3} we derive the $SU(2)$-equivariant dimensional reduction of Chern--Simons
gauge theories in arbitrary odd dimensionality and discuss some explicit examples.
In section \ref{sect4} we carry out the dimensional reduction of
five-dimensional Chern--Simons supergravity and point out some possible implications. Three appendices at the end of the paper contain some technical details which are used in the calculations of the main text: In appendix~\ref{app1} we summarise the pertinent group theory data for the classical gauge groups, in appendix~\ref{echfa} we explain the extended Cartan homotopy formula and some of its corollaries, and in appendix~\ref{sugrCS5} we describe the structure of Chern--Simons supergravity based on the supergroup $ SU(  2,2|N)$.

\newsection{Equivariant dimensional reduction} \label{sectionn2}

\subsection{Equivariant principal bundles}

In this paper we study gauge theories on the product
space $\mathcal{M}=M\times S^{2}$. Here $M$ is a closed
$d$-dimensional manifold with local coordinates $(x^\mu)_{\mu=1}^d$. On the sphere $S^{2}\simeq%
\mathbb{C}
P^{1}$ we use complex coordinates $(y,\bar y)$ defined by stereographic parameterization. 
We identify $S^{2}$ with the coset space $SU(2)/U(1)$. This induces a transitive action of $SU(2)$ on $S^{2}$ which we extend to the trivial action on $M$.
In order to obtain dimensionally reduced gauge invariant field theories starting from arbitrary gauge groups $\mathcal{G}$, in this section we study $SU(2)$-equivariant principal bundles on
$\mathcal{M}$ and their corresponding $SU(2)$-invariant connections. We follow for a large part the treatment of \cite{Manton:2010mj}.

Every $SU(2)$-equivariant principal bundle
over $S^{2}$ with structure group $\mathcal{G}$ is isomorphic to a quotient
space \cite{JSL}%
\begin{equation}
\mathcal{P}_{\rho}=SU(2)  \times_{\rho}\mathcal{G}\label{eq9}%
\end{equation}
where 
$\rho:U(1)  \rightarrow\mathcal{G}$ is a homomorphism and the elements of $SU(2)  \times_{\rho}\mathcal{G}$ are
equivalence classes $\left[s,g\right]$ on $SU(2)  \times\mathcal{G}$ with respect to the equivalence relation%
\begin{equation}
\left(  s,g\right)  \equiv\big(  s\, s_{0}\,,\,\rho( s_{0})
^{-1}\, g\big) \label{eq10}%
\end{equation}
for all elements $s_{0}\in U(1)\subset SU(2) $. The bundle projection
$\pi:\mathcal{P}_{\rho}\rightarrow S^{2}$ is given by%
\begin{equation}
\pi\left(  \left[  s,g\right]  \right)  =\left[  s\right] \label{eq11}%
\end{equation}
where $\left[  s\right]  $ denotes the left coset $
s\cdot U(1)   $ in $SU(2)  $. Bundles
$\mathcal{P}_{\rho},\mathcal{P}_{\rho'}$ are isomorphic if and only if
the homomorphisms $\rho,\rho'
:U(1)  \rightarrow\mathcal{G}$ take values in the same conjugacy class of $\mathcal{G}$.

Let $P$ be an $SU(2)$-equivariant principal $\CG$-bundle
over $\mathcal{M}=M\times S^{2}$ and select a good open covering $\left\{
U_{i}\right\}  _{i\in I}$ of $M$, i.e. all $U_{i}$ are contractible. Then
the restrictions $\left.  P\right\vert _{U_{i}\times S^{2}}$ are $SU(2)$-equivariant bundles which are trivial on each $U_{i}$, so
that %
\begin{equation}
\left.  P\right\vert _{U_{i}\times S^{2}}\simeq U_{i}\times \mathcal{P}_{\rho_{i}%
}\label{eq12}%
\end{equation}
where the homomorphisms $\rho_{i}:U(1)  \rightarrow
\mathcal{G}$ may be different for every open set $U_{i}\subset M$. However,
on the non-empty intersections $U_{ij}=U_{i}\cap U_{j}$ in $M$, the
restrictions $\left.  P\right\vert _{U_{ij}\times S^{2}}$ are isomorphic
to%
\begin{equation}
U_{ij}\times \mathcal{P}_{\rho_{j}}\simeq\left.  P\right\vert _{U_{ij}\times S^{2}%
}\simeq U_{ij}\times \mathcal{P}_{\rho_{i}} \ .\label{eq13}%
\end{equation}
This means that $\mathcal{P}_{\rho_{j}}\simeq \mathcal{P}_{\rho_{i}}$ and hence $\rho
_{i},\rho_{j}$ take values in the same conjugacy class of $\mathcal{G}$. If $M$ is connected, a
representative homomorphism $\rho$ can be chosen such that%
\begin{equation}
\left.  P\right\vert _{U_{i}\times S^{2}}\simeq U_{i}\times \mathcal{P}_{\rho
}\label{eq14}%
\end{equation}
for all $ i\in I$, and which satisfies%
\begin{equation}
\rho=h_{ij}^{-1}\, \rho \, h_{ij}\label{eq15}%
\end{equation}
for all transition functions $h_{ij}:U_{ij}\rightarrow\mathcal{G}$.
This implies that $h_{ij}$ take values in the centralizer of the image 
$\rho\left(  U(1)  \right)  $ in $\mathcal{G}$, which we denote by
\begin{equation}
\mathcal{H}=Z_{\mathcal{G}}\big(  \rho(  U(1)
)  \big) \ . \label{eq16}%
\end{equation}
Thus the collection of transition functions $\{h_{ij}\}$ for $ i,j\in I$ defines a principal bundle $P_{M}$ over $M$ with structure
group $\mathcal{H}$ which is the residual gauge group after dimensional reduction.

The homomorphism $\rho$ is determined by specifying a unique element $\Lambda
\in\mathfrak{g}$, where $\mathfrak{g}$ is the Lie algebra of $\mathcal{G}$. For this, introduce the Pauli spin matrices%
\begin{equation}%
\sigma_{1}=%
\begin{pmatrix}
0 & 1\\
1 & 0
\end{pmatrix}
\ , \qquad \sigma_{2}=%
\begin{pmatrix}
0 & -\ii\\
\ii & 0
\end{pmatrix}
\ , \qquad \sigma_{3}=%
\begin{pmatrix}
1 & 0\\
0 & -1
\end{pmatrix}
\label{eq17}%
\end{equation}
so that $T_{a}=-\frac{\ii}{2}\, \sigma_{a}$ for $a=1,2,3$ generate the defining representation of the Lie algebra $\mathfrak{su}(2)  $, where the $U(1)  $ subgroup of
$SU(2)  $ is generated by $T_{3}$. Any element of $U(1)  $ can be written as $\exp(  t\, T_{3})  $, where $t\in \mathbb{R}$, and the
image of this element under the homomorphism $\rho$ is 
\begin{equation}
\rho\big(  \exp( t\, T_{3})  \big)  =\exp( t\, \Lambda) \label{eq18}%
\end{equation}
where $\exp( t\, \Lambda )  \in\mathcal{G}$. 
Note that the identity element of $U(1)\subset SU(2)$ corresponds to $t=4\pi$, so that
\begin{equation}
\exp(  4\pi \, T_{3})  =\unit_{SU(2)} \ ,\label{eq19}%
\end{equation}
and since $\rho$ is a homomorphism 
it follows that $\Lambda$ must satisfy
\begin{equation}
\exp(  4\pi\, \Lambda)  =\unit_{\mathcal{G}} \ . \label{eq20}%
\end{equation}
This leads generally to an algebraic quantization condition on $\rho: U(1)\rightarrow \mathcal{G}$ which we describe explicitly in what follows.

The operations of restriction and induction \cite{AlGar12}
work for principal bundles in the same way as for
vector bundles. Given an $SU(2)$-equivariant principal bundle
$P\rightarrow M\times S^2 $, its restriction $P \vert _{M\times [\unit_{SU(2)}]}$ defines a $U(1)$-equivariant principal bundle on $M$ which is isomorphic to $P_{M}$. The $U(1)$-action on the fibre is defined by the homomorphism $\rho : U(1) \rightarrow \mathcal{G}$ and it extends trivially on the base space $M$. The inverse operation gives $P=SU(2)\times_{\rho}\left.P\right\vert _{M\times[\unit_{SU(2)}]  }$. 

\subsection{Invariant connections}

For a principal
bundle $P$ over $\mathcal{M}$ with connection one-form $\omega \in \Omega^{1}(P,\frg)$, a local gauge potential $\mathcal{A} \in \Omega^{1}(\CU,\frg)$ on a contractible open subset $\CU\subset \mathcal{M}$ is obtained
via a local section $\sigma:\CU\rightarrow P$ as the pull-back
\begin{equation}
\mathcal{A}=\sigma^{\ast}\omega \ .\label{2.1}%
\end{equation}
Let $\omega$ be an $SU(2)$-invariant connection on
$P \rightarrow  \mathcal{M}$. In each open subset $U_{i}\subset M$, we can pull-back $\omega$ to an $SU(2)$-invariant connection on $U_{i}\times
\mathcal{P}_{\rho}$ which corresponds to a gauge potential
$\mathcal{A}_{i}$ on $U_{i}\times S^{2}$ whose components are given by \cite{Manton:2010mj}%
\begin{align}
\mathcal{A}_{i,\mu} &  =A_{i,\mu} \ , \label{eq21}\\[4pt]
\mathcal{A}_{i,y} &  =\frac{-1}{1+y\, \bar{y}}\,\big(  \ii\bar{y}\, \Lambda+\Phi
_{i}\big) \ , \label{eq23}\\[4pt]
\mathcal{A}_{i,\bar{y}} &  =\frac{1}{1+y\,\bar{y}}\, \big(  \ii y\, \Lambda+\Phi
_{i}^{\dagger}\big) \ , \label{eq24}%
\end{align}
and are subjected to the invariance constraints
\begin{align}
\left[  \Lambda,A_{i,\mu}\right]   &  =0 \ , \label{eq25}\\[4pt]
\left[  \Lambda,\Phi_{i}\right]   &  =-\ii\Phi_{i} \ , \label{eq26}\\[4pt]
\big[  \Lambda,\Phi_{i}^{\dagger}\big]   &  =\ii\Phi_{i}^{\dagger
} \ . \label{eq27}%
\end{align}
On non-empty
overlaps $U_{ij}\subset M$ these fields obey the relations%
\begin{align}
{A}_{j} &  =h_{ij}^{-1}\, {A}_{i}\, h_{ij}+h_{ij}^{-1}%
\, \dd h_{ij} \ , \label{eq28}\\[4pt]
\Phi_{j} &  =h_{ij}^{-1}\, \Phi_{i}\, h_{ij} \ , \label{eq29}%
\end{align}
where $h_{ij}:U_{ij}\rightarrow\mathcal{H}$ are the transition functions of
$P_{M}$, and ${A}_{i}={A}_{i,\mu}\, \dd x^{\mu}$. The collection of
 local gauge potentials $A_{i}$ defines a connection on $P_{M}$, and
the constraints (\ref{eq25}) imply that $A_{i}$ take values in the Lie algebra $\mathfrak{h}$ of $\mathcal{H}$
which is consistent with $P_{M}$ having $\mathcal{H}$ as structure group. The
collection of local adjoint scalar fields $\Phi_{i}$ defines a section of the vector bundle ${\rm ad}(P_M):= P_{M}\times_{\text{ad}}\mathfrak{g}$ associated 
to $P_{M}$ by the adjoint representation of $\mathcal{H}$ on $\mathfrak{g}$. In the following we write $A$, $\Phi$ with $\left.  A\right\vert _{U_{i}}=A_{i}$ and $\left. \Phi\right\vert _{U_{i}}%
=\Phi_{i}.$

\subsection{Dimensional reduction of Yang--Mills theory}

On $\mathcal{M}=M\times S^{2}$ the metric
is taken to be the direct
product of a chosen metric $g_{\mu\nu}$ on $M$ and the round metric of the two-sphere, so that
\begin{equation}
\dd s^{2}
=G_{\mu^{\prime}\nu^{\prime}}\,\dd x^{\mu^{\prime}}\otimes \dd x^{\nu^{\prime}}
 =g_{\mu\nu}\, \dd x^{\mu} \otimes \dd x^{\nu}+\frac{4R^{2}}{\left(  1+y\, \bar{y}\right)
^{2}}\, \dd y \otimes \dd\bar{y}\label{ym5}%
\end{equation}
where the indices $\mu^{\prime},\nu^{\prime}$ run over $1, \ldots,  d+2$ and $R$ is the radius of $S^2$. For a principal $\mathcal{G}$-bundle $P \rightarrow \mathcal{M}$ with gauge potential $\alg$, the 
 Yang--Mills Lagrangian is given by
\begin{equation}
L_{\rm{YM}}=-\frac{1}{4g_{\rm YM}^2}\, \sqrt{G}~ \text{\textsf{Tr}}\big( \mathcal{F}_{\mu^{\prime}%
\nu^{\prime}}\, \mathcal{F}^{\mu^{\prime}\nu^{\prime}} \big) \label{nym1}%
\end{equation}
where $\mathcal{F}$ is the curvature two-form
\begin{equation}
\mathcal{F} 
=\dd\mathcal{A+A\wedge A}=\mbox{$\frac{1}{2}$}\, \mathcal{F}_{\mu^{\prime}\nu^{\prime}}\, \dd x^{\mu^{\prime}}\wedge \dd x^{\nu^{\prime}}\label{nym2}
\end{equation}
and $G= \det (G_{\mu^{\prime}\nu^{\prime}} )$. Here $g_{\rm YM}$ is the Yang--Mills coupling constant and $\textsf{Tr}$ denotes a non-degenerate invariant quadratic form on the Lie algebra $\frg$ of the gauge group $\mathcal{G}$, which for $\mathcal{G}$ semisimple is proportional to the Killing--Cartan form.

Expanding (\ref{nym1}) into components along $M$ and $\mathbb{C}P^{1}$ we get%
\begin{equation}
L_{\mathrm{YM}}=-\frac{1}{4g_{\rm YM}^2}\, \sqrt{G}~\text{\textsf{Tr}}\Big( \mathcal{F}_{\mu\nu
}\, \mathcal{F}^{\mu\nu}+\frac{\left(  1+y\, \bar{y}\right)  ^{2}}{2R^{2}}\, g^{\mu\nu
}\, \big(  \mathcal{F}_{\mu y}\, \mathcal{F}_{\nu\bar{y}}+\mathcal{F}_{\mu\bar{y}%
}\, \mathcal{F}_{\nu y}\big)  +\frac{\left(  1+y\, \bar{y}\right)  ^{4}}{8R^{4}%
}\, \mathcal{F}_{y\bar{y}}\, \mathcal{F}_{\bar{y}y}\Big)
\end{equation}
where from (\ref{eq21})--(\ref{eq24}) we have
\begin{align}
\mathcal{F}_{\mu\nu} &  =F_{\mu\nu} \ , \label{ym22}\\[4pt]
\mathcal{F}_{\mu y} &  =-\frac{1}{1+y\, \bar{y}}\, \nabla_{\mu}\Phi \ , \label{ym23}\\[4pt]
\mathcal{F}_{\mu\bar{y}} &  =\frac{1}{1+y\, \bar{y}}\, \nabla_{\mu}\Phi^{\dagger
} \ , \label{ym24}\\[4pt]
\mathcal{F}_{y\bar{y}} &  =\frac{1}{\left(  1+y\, \bar{y}\right)  ^{2}}\, \big(
2\ii\Lambda-\big[  \Phi,\Phi^{\dagger}\big] \big) \ , \label{ym27}%
\end{align} 
with
\begin{align}
F  & =\dd A+A\wedge A=\mbox{$\frac{1}{2}$}\, F_{\mu\nu}\, \dd x^{\mu}\wedge \dd x^{\nu} \ , \label{curv}\\[4pt]
\nabla\Phi & =\dd\Phi+\left[  A,\Phi\right] = \nabla_\mu\Phi\, \dd x^\mu \label{covd} \ .
\end{align}
Integrating the corresponding Yang--Mills action 
\begin{equation}
S_{\mathrm{YM}}=\int_{\mathcal{M}}\, \dd^{d+2}x\ \sqrt{G}\ L_{\mathrm{YM}}
\end{equation}
over $S^2 \simeq \mathbb{C}P^{1}$ using
\begin{equation}
\int_{\mathbb{C}P^{1}}\, \frac{R^{2}}{\left(  1+y\, \bar{y}\right)
  ^{2}}\, \dd y \wedge \dd\bar{y}=4\pi \, R^{2} \ ,
\end{equation}
we get the action
\begin{align}
S_{\mathrm{YMH}}=\frac{\pi\, R^{2}}{g_{\rm YM}^{2}}\, \int_{M}\, \dd^{d}%
x~\sqrt{g}\ \text{\textsf{Tr}}\Big( F_{\mu\nu}\, \left(  F^{\mu\nu}\right)  ^{\dagger
}& +\frac{1}{2R^{2}}\, \big(  \nabla_{\mu}\Phi \, \nabla^{\mu}\Phi^{\dagger}+\nabla_{\mu}%
\Phi^{\dagger}\, \nabla^{\mu}\Phi\big) \nonumber \\ 
& + \frac{1}{8R^{4}}\, \big(  2\ii\Lambda-\big[  \Phi,\Phi^{\dagger}\big]
\big)  ^{2}\Big) \label{ymredact}
\end{align}
which describes a Yang--Mills--Higgs theory on $M$ with gauge group $\mathcal{H}$ \cite{ForgaCS:1979zs,Dolan:2009ie,Manton:2010mj}.

\newsection{Principal quiver bundles} \label{sector3}

In order to solve the constraint equations (\ref{eq25})--(\ref{eq27}) explicitly, it is necessary to fix the element
$\Lambda\in\frg$ and therefore the gauge group $\mathcal{G}$. In this section we consider the case where $\CG$ is one of the classical Lie groups $U(n)$, $SO(2n)$,
$SO(2n+1)$, or $Sp(2n)$. In this case equivariant dimensional reduction gives principal $\mathcal{H}$-bundles $P_{M} \rightarrow M$ which can be characterized in terms of quivers, and (\ref{ymredact}) becomes an action for a quiver gauge theory on $M$.

In
the Cartan--Weyl basis, the generators of the gauge group $\mathcal{G}$ satisfy the
commutation relations%
\begin{align}
\left[  H_{i},H_{j}\right]   &  =0 \ , \label{cwb1}\\[4pt]
\left[  H_{i},X_{\alpha}\right]   &  =\alpha_{i}\, X_{\alpha} \ , \label{cwb2}\\[4pt]
\left[  X_{\alpha},X_{\beta}\right]   &  =\left\{
\begin{array}
[c]{l}%
N_{\alpha,\beta}\, X_{\alpha+\beta} \quad \text{ ~~if }\alpha+\beta\text{ is a root} \ , \\
0 \quad \text{ ~~~~~~~~~~~~~~~otherwise} \ ,
\end{array}
\right.  \text{ }\label{cwb3}\\[4pt]
\left[  X_{\alpha},X_{-\alpha}\right]   &  =\frac{2}{\left\vert \alpha
\right\vert ^{2}}\, \sum\limits_{i=1}^{n}\,\alpha_{i}\, H_{i}\label{cwb4} \ ,
\end{align}
where $n$ is the rank of $\mathcal{G}$, the subset $\left\{  {H}_{i}\right\}_{i=1}^n  $ generates the Cartan subalgebra
$\mathfrak{t\subset g}$, the vectors $\alpha$ are the roots of the Lie algebra $\mathfrak{g}$ of $\mathcal{G}$, and $\left\{
X_{\alpha}\right\}  $ are the root vectors with normalization constants $N_{\alpha,\beta}$. By gauge invariance, the element $\Lambda
\in\mathfrak{g}$ can be conjugated into the Cartan subalgebra
generated by $\left\{H_{i}\right\}  $. Then there is still a residual gauge symmetry under the discrete Weyl subgroup $\Wcal\subset\mathcal{G}$ which acts by permuting the eigenvalues $\lambda_{i}$, $i=1,\dots, n$ of $\Lambda$. We can use this symmetry to group $\lambda_{i}$ into $m+1$ degenerate blocks, $0\leq m\leq n-1$, of dimensions $k_{\ell}$ such that $\lambda_{k_0+ k_{1}+\cdots+k_{\ell-1}+1}=\cdots=\lambda_{k_0+ k_{1}+\cdots+k_{\ell-1}+k_{\ell}}=:\alpha_{\ell}$ for $\ell=0, 1,\ldots,m$, where $k_{-1}:=0$ and
\begin{equation}
\sum\limits_{\ell=0}^{m}\, k_{\ell}=n \ . \label{cb1}%
\end{equation}
Then the element $\Lambda$ can be expanded as
\begin{equation}
\Lambda=\ii\sum\limits_{\ell=0}^{m}\, \alpha_{\ell} \ \sum\limits_{i=1}^{k_{\ell}} \,
H_{k_{1}+\cdots+k_{\ell-1}+i} \ . \label{cwb5}%
\end{equation}
Similarly, the Higgs fields $\Phi$ and the gauge field $A$ can both be
expanded in the Cartan--Weyl basis as
\begin{align}
\Phi&=\sum\limits_{i=1}^{n}\, \phi_{i}\, H_{i}+\sum\limits_{\alpha>0}\, \big(
\phi_{\alpha}\, X_{\alpha}+\phi_{-\alpha}\, X_{-\alpha}\, \big) \ ,
\label{cw6} \\[4pt]
A&=\sum\limits_{i=1}^{n}\, A_{i}\, H_{i}+\sum\limits_{\alpha>0}\, \big(
A_{\alpha}\, X_{\alpha}+A_{-\alpha}\, X_{-\alpha}\big) \ .
\label{cb7}%
\end{align}

Let us first consider the unitary gauge group $\mathcal{G}=U(n)
$. Since $\Lambda\in\mathfrak{u}(n)$, it may be represented by a
Hermitian $n\times n$ matrix which can always be taken to be diagonal
by conjugation with a suitable element $g \in U(n)  $. The roots and
the forms of the generators in the Cartan--Weyl basis are summarized in appendix \ref{app1}.

Using
\begin{equation}
\left[  H_{i},X_{e_{j}-e_{k}}\right]  =\left(
\delta_{ji}-\delta_{ki}\right) \, X_{e_{j}-e_{k}}\label{u3}
\end{equation}
the invariance constraints (\ref{eq26}) and (\ref{eq27}) yield
\begin{equation}
\phi_{i}=0 \ , \qquad \phi_{jk}\, \left(
  \lambda_{j}-\lambda_{k}+1\right)    =0 = \phi_{kj}\, \left(
  \lambda_{k}-\lambda_{j}+1\right) \ . \label{u7} 
\end{equation}
To allow for non-trivial solutions, it is necessary to require
$\lambda_{k}-\lambda_{j}=\pm\, 1$. Using Weyl symmetry to restrict attention to
$
\lambda_{j}-\lambda_{k}=-1\label{er44}%
$
with $\lambda_{j}\neq\lambda_{k}\neq0$, we find $\phi_{kj}=0$ while $\phi_{jk} $ can be non-vanishing.
However, not all of the fields $\phi_{jk}$ are non-zero.
The only non-vanishing components arise when $j$ and $k
$ belong to neighbouring blocks of indices. 
If $j,k$ belong to the same block $K_{\left(\ell\right)  }:=\{{k_0+ k_{1}+\cdots +k_{\ell-1}+i }\}_{i=1}^{k_{\ell}}$, then $\lambda_{j}=\lambda_{k}=\alpha_{\ell}$ and so $\phi_{jk}=0$ by (\ref{u7}).
On the other hand, if  $j\in K_{\left(  \ell\right)  }$ and  $k\in K_{\left(  \ell+1\right)  }$, then $\lambda_{j}=\alpha_{\ell}$ and $\lambda_{k}=\alpha_{\ell+1}$, and by (\ref{u7}) if $\phi_{jk}\neq0$ then $\alpha_{\ell}-\alpha_{\ell+1}=-1$, so we have 
$
\alpha_{\ell}  =\alpha+\ell
$
for $\ell=0, 1, \ldots,m$ and $\alpha :=\alpha_{0}$. Therefore the Higgs field (\ref{cw6}) has the form
\begin{equation}
\Phi=\sum\limits_{\ell=0}^{m}\, \phi_{(\ell+1)}\label{u19}%
\end{equation}
where%
\begin{equation}
\phi_{(\ell+1)}=\sum\limits_{\stackrel{\scriptstyle j\in K_{\left(  \ell\right)  }\,,\, k\in K_{\left(
\ell+1\right)  }}{\scriptstyle j<k}} \, \phi_{jk} \, X_{e_{j}-e_{k}}
\end{equation}
with $\phi_{(m+1)}:=0$.

The constraint equation (\ref{eq25}) gives
\begin{equation}
A_{jk}\, \left(  \lambda_{j}-\lambda_{k}\right)  =0=
A_{kj}\, \left(  \lambda_{k}-\lambda_{j}\right) \ . \label{u22}%
\end{equation}
Here non-trivial solutions occur when $\lambda_{k}=\lambda_{j}$. This
happens when $j,k$ belong to the same block $K_{\left(  \ell\right)  }$ and thus%
\begin{equation}
A=\sum\limits_{\ell=0}^{m}\, A_{(\ell)}\label{u23}%
\end{equation}
where%
\begin{equation}
A_{(\ell)}=\sum\limits_{i\in K_{\left(  \ell\right)  }}\, A_{i}\, H_{i}%
+\sum\limits_{\stackrel{\scriptstyle j,k\in K_{\left(  \ell\right)  }}{\scriptstyle j<k}}\, \left(
A_{jk}\, X_{e_{j}-e_{k}}+A_{kj}\, X_{e_{k}-e_{j}}\right) \ .
\end{equation}
This calculation also shows that the breaking of the original $U(n)$ gauge symmetry to the centralizer subgroup (\ref{eq16}) is given by
\begin{equation}
\mathcal{H}=\prod\limits_{\ell=0}^{m}\, U(k_{\ell}) \ . \label{u24}%
\end{equation}

The $\mathfrak{u}(n)$-valued gauge potential $\mathcal{A}$ on $\mathcal{M}$ is by construction $SU(2)$-invariant and 
decomposes into $k_{\ell}\times k_{\ell'}$ blocks $\mathcal{A}^{\ell,\ell'}$
with $\ell,\ell'=0, 1, \ldots ,m$ and
\begin{align}
\mathcal{A}^{\ell,\ell} &  =A_{(\ell)}-\a_{(\ell)} \ , \label{ansa1}\\[4pt]
\mathcal{A}^{\ell, \ell+1} &  =-\phi_{(\ell+1)}\, \beta \ , \label{ansa2}\\[4pt]
\mathcal{A}^{\ell+1, \ell} &  =-\big(  \mathcal{A}^{\ell, \ell+1}\big)  ^{\dagger}%
=\phi_{(\ell+1)}^{\dagger}\, \bar{\beta} \ , \\[4pt]
\mathcal{A}^{\ell+i, \ell} &  =0=\mathcal{A}^{\ell, \ell+i} \qquad \text{for} \quad i\geq2 \ . \label{ansa4}
\end{align}
Here the local one-forms $\a_{(\ell)}$ on $\C P^1$ are given by
\begin{equation}
\a_{(\ell)}=-\frac{\alpha_{\ell}\, \left(  \bar{y}\, \dd y-y\, \dd\bar{y}\right) }{ 1+y\, \bar
{y}} \ , \label{magm}
\end{equation} 
and 
\begin{equation}
\beta=\frac{\dd y}{1+y\, \bar{y}} \ , \qquad \bar{\beta}=\frac
{\dd\bar{y}}{1+y\, \bar{y}} \label{cdf}
\end{equation}
are the unique covariantly constant $SU(2)$-invariant $(1,0)$- and $(0,1)$-forms on $\mathbb{C}P^1$ respectively. From (\ref{ansa1})--(\ref{ansa4}) it follows that the curvature two-form splits into $k_{\ell}\times k_{\ell'}$ blocks 
\begin{equation}
\mathcal{F}^{\ell,\ell'}=\dd\mathcal{A}^{\ell,\ell'}+\sum\limits_{\ell''=0}^{m}\, \mathcal{A}%
^{\ell,\ell''} \wedge \mathcal{A}^{\ell'',\ell'}\label{lolazo}
\end{equation} 
and its only non-vanishing components are
\begin{align}
\mathcal{F}^{\ell,\ell} &  =F_{(\ell)}-\sff_{(\ell)}+\big(  \phi_{(\ell)}^{\dagger}\, \phi_{(\ell)}-\phi
_{(\ell+1)}\, \phi_{(\ell+1)}^{\dagger}\big) \, \beta \wedge \bar{\beta} \ , \nonumber \\[4pt]
\mathcal{F}^{\ell,\ell+1} &  =-\nabla\phi_{(\ell+1)}\wedge \beta \ , \nonumber \\[4pt]
\mathcal{F}^{\ell+1,\ell} &  =\nabla\phi_{(\ell+1)}^{\dagger}\wedge
\bar{\beta} \ ,
\end{align}
where
\begin{align}
\sff_{(\ell)} &  =2\alpha_\ell\, \beta \wedge \bar{\beta} \ , \nonumber \\[4pt]
F_{(\ell)}& = \dd A_{(\ell)}+A_{(\ell)}\wedge A_{(\ell)} \ , \nonumber \\[4pt]
 \nabla\phi_{(\ell+1)} &  =\dd\phi_{(\ell+1)}+A_{(\ell)}\, \phi_{(\ell+1)}-\phi_{(\ell+1)}\, A_{(\ell+1)} \ , \nonumber \\[4pt]
\nabla\phi_{(\ell+1)}^{\dagger} &  =\dd\phi_{(\ell+1)}^{\dagger}+A_{(\ell+1)}\, \phi_{(\ell+1)}^{\dagger
}-\phi_{(\ell+1)}^{\dagger}\, A_{(\ell)}
\end{align}
with $\phi_{(0)}:=0=:\phi_{(m+1)}$.

The eigenvalues of the matrix $\Lambda$ from (\ref{cwb5}) are constrained by (\ref{eq20}) to quantized values $\alpha_{\ell} \in \frac12\, \mathbb{Z}$ given by
\begin{equation}
\alpha_{\ell}=\frac{p+2\ell}{2}
\end{equation}
for arbitrary $p\in\Z$. 
It follows that the matrix $\Lambda$ geometrically parameterizes the quantized magnetic charges of the unique $SU(2)$-invariant family of monopole connections $\a_{(\ell)}$ on $\C P^1$. With $p=-m$ the Yang--Mills--Higgs model (\ref{ymredact}) reproduces the quiver gauge theories from~\cite{Popov:2005ik} which are based on the linear $A_{m}$ quivers
\begin{equation}
\xymatrix@R=0.28pc{
\bullet \ar[r] & \bullet \ar[r]&\bullet \ \cdots \ \bullet \ar[r]& \bullet 
} \label{quiver}
\end{equation}
containing $m+1$ nodes corresponding to the gauge groups $U(k_\ell)$ and gauge fields $A_{(\ell)}$, and $m$ arrows corresponding to the $U(k_{\ell+1})\times U(k_\ell)$ bifundamental Higgs fields $\phi_{(\ell+1)}$. The quiver (\ref{quiver}) characterizes how $SU(2)$-invariance is incorporated into the gauge theory on $\CM=M\times S^2$.

Note that this correspondence with quivers is somewhat symbolic, as an
$SU(2)$-equivariant principal $\CG$-bundle does not belong to a
representation category for the quiver (\ref{quiver}). The association
is possible because in the present case the gauge group $\CG$ is a
matrix Lie group: One may regard $U(k_\ell)$ as the group of unitary
automorphisms of a complex inner product space $V_{k_\ell}\simeq
\C^{k_\ell}$ and the Higgs fields $\phi_{(\ell+1)}$ fibrewise as maps
in $\Hom(V_{k_{\ell+1}}, V_{k_\ell})$. To associate a quiver bundle to
our construction we need a suitable representation of the quiver
(\ref{quiver}) in the category of vector bundles on $M$. For this, we
can take the complex vector bundle $E=P\times_\varrho V$ on $\CM$
associated to the fundamental representation $\varrho: \CG\to U(V)$ of
$\CG=U(n)$ on $V\simeq \C^n$. Then the restriction $E_M:= E|_{M\times[\unit_{SU(2)}]}=P_M\times_\varrho V|_\CH$ is a $U(1)$-equivariant vector bundle on $M$ with fibre the restriction $V|_\CH=\bigoplus_{\ell=0}^m\, V_{k_\ell}$ of the linear representation $(\varrho,V)$ to $\CH$. The $U(1)$-action on the fibre is given by $\exp(t\, \Lambda)|_{V_{k_\ell}}=\e^{\ii t\, (\frac p2+\ell)}\, \unit_{V_{k_\ell}}$ and the Higgs fields are morphisms $\Phi|_{E_{k_{\ell+1}}}:E_{k_{\ell+1}} \to E_{k_\ell}$ of the vector bundles $E_{k_\ell}:=P_M\times_\varrho V_{k_\ell}$ for each $\ell=0,1,\dots,m$.

Our detailed treatment here of the standard case with $\CG=U(n)$ has
the virtue that the exact same analysis can be performed for the
remaining classical gauge groups $\mathcal{G}=SO(2n)$,  $SO(2n+1)$,
and $Sp(2n)$; the requisite group theory data for their decompositions
in the Cartan--Weyl basis are summarised in appendix~\ref{app1}. In
every case one shows that, for generic eigenvalues $\alpha_\ell$ of
the matrix $\Lambda$, the residual gauge symmetry group is again given
by (\ref{u24}) (as a subgroup of $\CG$) and the structure of the
dimensionally reduced gauge theory can again be encoded in the $A_m$
quiver (\ref{quiver}), with only trivial redefinitions of the coupling
constants in (\ref{ymredact}) distinguishing the different cases. Such
redefinitions may have implications in matching the quiver gauge
theories with more realistic models as in~\cite{Dolan:2009ie}.

\newsection{Covariant quiver gauge theories} \label{sect3}

\subsection{Chern--Simons theory and transgression forms}

Let ${P}$ be a principal
bundle with $\left(  2n+1\right)  $-dimensional base space $\mathcal{M}$ and
structure group $\mathcal{G}$. Let $\mathfrak{g}$ be the Lie algebra of $\mathcal{G}$ generated by $ \mathsf{T}_{a}$
with $a=1,\ldots,\dim\mathfrak{g}$. Let $\mathcal{A}$ be a gauge potential on
$\mathcal{M}$ defined as in (\ref{2.1}) with
\begin{equation}
\mathcal{A=A}_{\mu^{\prime}} \, \dd x^{\mu^{\prime}} =\mathcal{A}_{\mu^{\prime}}^{a} \, \dd x^{\mu^{\prime}} \otimes \mathsf{T}_{a}%
\end{equation}
and let $\mathcal{F}$ be the curvature two-form (\ref{nym2}).
The product%
\begin{equation}
\chi^{\left(  2n+2\right)  }(\alg) =\big\langle \mathcal{F}^{n+1}\big\rangle
\end{equation}
is a closed $\left(2n+2\right)$-form on $\mathcal{M}$ which defines a characteristic
class of ${P}$, where the
bracket 
\begin{equation}
\left\langle - \right\rangle \,:\, \mathfrak{g}^{\otimes(n+1)} \ \longrightarrow \ 
\mathbb{R}
\end{equation}
denotes a symmetric $\mathfrak{g}$-invariant polynomial of rank $n+1$ which can always be determined once an explicit presentation of
$\mathfrak{g}$ is specified.
Since $\chi^{(2n+2)}(\alg)$ is closed, it
can be locally written as the exterior derivative of a Chern--Simons form%
\begin{equation}
\chi^{\left(  2n+2\right)  }(\alg) =\dd{L}_{\mathrm{CS}}^{\left(  2n+1\right)  }
\end{equation}
where%
\begin{equation}
{L}_{\mathrm{CS}}^{\left(  2n+1\right)  } =\left(  n+1\right) \, \int_{0}%
^{1}\, \dd t\ \left\langle \mathcal{A}\wedge \left(  t\,\dd\mathcal{A}%
+t^{2}\,\mathcal{A}\wedge \mathcal{A}\right)  ^{n}\right\rangle \ .
\label{CSgfgf}%
\end{equation}
The Chern--Simons form is, up to boundary terms, gauge-invariant. This means
that under infinitesimal gauge transformations with parameter function $\lambda= \lambda^a\otimes \mathsf{T}_a\in\Omega^0(\CM,\frg)$, the variation of the 
gauge potential is given by
\begin{equation}
\delta_\lambda\mathcal{A}=\dd\lambda+ \left[  \mathcal{A},\lambda
\right]
\end{equation}
and subsequently (\ref{CSgfgf}) remains unchanged modulo
boundary terms. Due to their quasi-gauge invariance property, Chern--Simons forms are
good candidates to construct action functionals with the gauge potential
$\mathcal{A}$ as the fundamental field, so we define
\begin{equation}
{S}_{\mathrm{CS}}^{\left(  2n+1\right)  } =\kappa\, \left(  n+1\right) \, \int_{\mathcal{M}%
} \ \int_{0}^{1}\, \dd t\ \left\langle \mathcal{A}\wedge\left(  t\, \dd\mathcal{A}%
+t^{2}\, \mathcal{A}\wedge \mathcal{A}\right)  ^{n}\right\rangle \label{CSact}
\end{equation}
where $\kappa$ is a coupling constant.

Chern--Simons forms are a particular case of more general globally-defined differential forms. Consider the action functional
\begin{equation}
{S}_{{T}}^{\left(  2n+1\right)  }\left[  \mathcal{A},\mathcal{\bar{A}}\, \right]
=\kappa\, \int_{\mathcal{M}}\, Q_{\mathcal{A}\leftarrow\mathcal{\bar{A}}}^{\left(
2n+1\right)  }\label{tfcvcv}%
\end{equation}
where $Q_{\mathcal{A}\leftarrow
\mathcal{\bar{A}}}^{\left(  2n+1\right)  }$ is the
\textit{transgression form} \cite{Nakahara,Azcarraga,Borowiec:2003rd,Mora:2006ka} defined by%
\begin{equation}
Q_{\mathcal{A}\leftarrow\mathcal{\bar{A}}}^{\left(  2n+1\right)
}=- Q_{\mathcal{\bar{A}}\leftarrow\mathcal{A}}^{\left(  2n+1\right)  }
:= \left(
n+1\right) \, \int_{0}^{1}\, \dd t\ \left\langle \left(  \mathcal{A}-\mathcal{\bar{A}%
}\, \right)  \wedge\mathcal{F}_{t}^{n}\right\rangle \ .
\end{equation}
Here $\mathcal{A}$ and $\mathcal{\bar{A}}$ are two $\frg$-valued gauge potentials
and we set
\begin{equation}
\mathcal{A}_{t} =\mathcal{\bar{A}}+t\, \left(  \mathcal{A}-\mathcal{\bar{A}%
}\, \right) \ , \qquad \mathcal{F}_{t}   =\dd\mathcal{A}_{t}+\mathcal{A}_{t}\wedge\mathcal{A}_{t} \ .
\end{equation}
The action (\ref{tfcvcv}) is invariant under two
different sets of symmetries. On the one hand, it is diffeomorphism
invariant since it is constructed using only differential forms on $\CM$, while on the other
hand it possesses full invariance under local gauge transformations \cite{Izaurieta:2005vp}
\begin{equation}
\mathcal{A}^{g} =g^{-1}\, \mathcal{A}\, g+g^{-1}\, \dd g \ , \qquad
\mathcal{\bar{A}}^{g}  =g^{-1}\, \mathcal{\bar{A}}\, g+g^{-1}\, \dd g
\label{calAgaugetransf}\end{equation}
where $g =\exp\left(  \lambda^{a}
\otimes \mathsf{T}_{a}\right) \in \Omega^0(\CM,\mathcal{G} ) $. The Euler--Lagrange field equations associated to (\ref{tfcvcv}) read as
\begin{equation}
\big\langle \mathcal{F}^{n}\, \mathsf{T}_{a}\big\rangle = 0=
\big\langle \mathcal{\bar{F}}^{n} \, \mathsf{T}_{a}\big\rangle
\label{calFfieldeq}\end{equation}
for $a=1,\dots,\dim\frg$, subject to the boundary conditions 
\begin{equation}
\int_{0}^{1}\, \dd t\ \left\langle \delta\mathcal{A}_{t}\wedge\left(
\mathcal{A}-\mathcal{\bar{A}}\, \right)
\wedge\mathcal{F}^{n-1}\right\rangle \, 
\Big\vert _{\partial\mathcal{M}}
=0%
\end{equation}
for arbitrary variations $\delta\alg_t$ of the gauge potentials. It is easy to check that the Chern--Simons form is recovered in the limit
$\mathcal{\bar{A}}=0$.

\subsection{Topological Chern--Simons--Higgs models}

We will now perform the $SU(2)$-equivariant dimensional
reduction of the Chern--Simons gauge theory on $\mathcal{M}=M \times
S^2$, where $M$ is an oriented manifold of dimension $d=2n-1$. Throughout we assume that the manifold $M$ is closed, as no novel boundary effects arise in the models we derive.
The gauge field defined by (\ref{eq21})--(\ref{eq24}) can be written in the form
\begin{equation}
\mathcal{A}=A-\a-\Phi\otimes \beta+\Phi^{\dagger
}\otimes \bar{\beta} \ ,
\end{equation}
where
\begin{equation}
\a:=\Lambda\otimes\frac{\ii\left(  \bar{y}\, \dd y-y\, \dd\bar{y}\right) }{ 1+y\, \bar
{y}} 
\end{equation}
and we have used (\ref{cdf}). 
In general, the computation of the reduced Chern--Simons action directly from its definition (\ref{CSact}) is somewhat involved; to simplify the calculations considerably we use the subspace separation method~\cite{Izaurieta:2006wv} which provides a systematic way
to compute Chern--Simons forms by making use of the extended Cartan homotopy
formula \cite{Manes:2012zz} (see appendix~\ref{echfa}). This method has the virtue of separating the action into bulk and boundary contributions, and also splitting the Lagrangian into terms valued in a specified subspace decomposition of the gauge
algebra.

The applicability of the method relies on regarding Chern--Simons forms as
transgression forms that satisfy the \emph{triangle equation} (see appendix \ref{triangleq})
\begin{equation}
Q_{\mathcal{A}_{2}\leftarrow\mathcal{A}_{0}}^{\left(  2n+1\right)
}=Q_{\mathcal{A}_{2}\leftarrow\mathcal{A}_{1}}^{\left(  2n+1\right)
}+Q_{\mathcal{A}_{1}\leftarrow\mathcal{A}_{0}}^{\left(  2n+1\right)
}+\dd Q_{\mathcal{A}_{2}\leftarrow\mathcal{A}_{1}\leftarrow\mathcal{A}_{0}%
}^{\left(  2n\right)  }  \ ,\label{trieq}%
\end{equation}
which decomposes a transgression form into the sum of two transgression forms
depending on an intermediate connection plus an exact form with
\begin{equation}
Q_{\mathcal{A}_{2}\leftarrow\mathcal{A}_{1}\leftarrow\mathcal{A}_{0}}^{\left(
2n\right)  }:= n\, \left(  n+1\right) \, \int_{0}^{1}\, \dd t \ \int_{0}^{t}\,
\dd s\ \left\langle \left(  \mathcal{A}_{2}-\mathcal{A}_{1}\right)  \wedge\left(
\mathcal{A}_{1}-\mathcal{A}_{0}\right)  \wedge\mathcal{F}_{s,t}^{n-1}%
\right\rangle
\label{Q2n}\end{equation}
where $\mathcal{F}_{s,t}  =\dd\mathcal{A}_{s,t}+\mathcal{A}_{s,t}\wedge\mathcal{A}%
_{s,t}$ with $\mathcal{A}_{s,t}   =s\, \left(  \mathcal{A}_{2}-\mathcal{A}_{1}\right)
+t\, \left(  \mathcal{A}_{1}-\mathcal{A}_{0}\right)  +\mathcal{A}_{0}.$
The method then proceeds in three steps:
\begin{enumerate}
\item Decompose the gauge algebra $\mathfrak{g}$ into $p+1$ vector subspaces
$\mathfrak{g}=\frg_{0}\oplus \frg_1\oplus \cdots\oplus \frg_{p}$.
\item Expand the connections into components valued in each subspace as $\mathcal{A}=\mbf{a}_{0}+\mbf{a}_1+\cdots+\mbf{a}_{p}$ and $\mathcal{\bar{A}}=\bar{\mbf{{a}}}_{0}+\bar{\mbf{a}}_1+ \cdots +\bar{\mbf{{a}}}_{p}$ with $\mbf{a}_i,\bar{\mbf{a}}_i\in\frg_i$ for $i=0,1,\ldots,p$.
\item Evaluate the triangle equation (\ref{trieq}) with the connections written in terms of
pieces valued in each subspace as
\begin{equation}
\mathcal{A}_{0}  =\mathcal{\bar{A}} \ ,\qquad
\mathcal{A}_{1}   =\mbf{a}_{0}+\mbf{a}_1+\cdots +\mbf{a}_{p-1} \ ,\qquad
\mathcal{A}_{2}  =\mathcal{A} \ .%
\end{equation}
\item Repeat step 3 for the transgression form $Q_{\mathcal{A}_{1}\leftarrow
\mathcal{A}_{0}}^{\left(  2n+1\right)  }$, and so on.
\end{enumerate}

For the present case we decompose $\frg=\frg_0\oplus\frg_1$ with $\frg_0=\frh$ and $\frg_1= \frg\ominus\frh$, and expand the gauge potential as
\begin{align}
\mathcal{A}_{0} &  =0 \ ,\label{ssmd1}\\[4pt]
\mathcal{A}_{1} &  =-\a \ ,\label{ssmd2}\\[4pt]
\mathcal{A}_{2} &  =A-\a \ ,\label{ssmd3}\\[4pt]
\mathcal{A}_{3} &  =\Phi^{\dagger}\otimes \bar\beta-\Phi\otimes
\beta+A-\a \ . \label{ssmd4}%
\end{align}
By applying the triangle equation (\ref{trieq}) we obtain the 
expression for the reduced Chern--Simons action: The reduced Lagrangian splits into the sum of three terms%
\begin{align}
{L}_{\Phi} &  =\kappa\, Q_{\mathcal{A}_{3}\leftarrow\mathcal{A}_{2}}^{\left(
2n+1\right)  }=2\kappa\, \left(  n+1\right) \, \int_{0}^{1}\,\dd t\
\left\langle t\, \big(  \Phi
\, \nabla\Phi^{\dagger}-\Phi^{\dagger}\, \nabla\Phi\big) \wedge \beta \wedge\bar{\beta
} \wedge F^{n-1}\right\rangle \ , \nonumber \\[4pt]
{L}_{A} &  =\kappa \, Q_{\mathcal{A}_{2}\leftarrow\mathcal{A}_{1}}^{\left(
2n+1\right)  }=2\kappa\, \left(  n+1\right)\,  \int_{0}^{1}\, \dd t\
\left\langle 2\ii \Lambda\, \beta
\wedge \bar{\beta} \wedge
A\wedge\big(  t\,\dd A+t^{2}\, A\wedge A\big)  ^{n-1}\right\rangle \ , \nonumber \\[4pt]
{L}_{\Lambda} &  =\kappa\, Q_{\mathcal{A}_{1}\leftarrow\mathcal{A}_{0}}^{\left(
2n+1\right)  }=0 \ .
\end{align}
By integrating over $S^2$, the original $(2n+1)$-dimensional Chern--Simons gauge theory reduces to a Chern--Simons--Higgs type model in $d=2n-1$ dimensions with action
\begin{equation}
{S}_{\mathrm{CSH}}^{\left(  2n-1\right)  }=  \kappa^{\prime}\, 
\int_{M} \ \int_{0}^{1}\, \dd t\ \left\langle t\, \big(  \Phi\, \nabla\Phi^{\dagger} 
-\Phi^{\dagger}\, \nabla\Phi\big)  \wedge F^{n-1} +2\ii\Lambda\, A\wedge \big(
t\,\dd A+t^{2}\, A\wedge A\big)  ^{n-1}\right\rangle \label{CSredact}%
\end{equation}
subject to the constraints (\ref{eq25})--(\ref{eq27}).
Here we have defined $\kappa^{\prime}=8\pi\, R^{2}\,\left(  n+1\right)
\, \kappa$ and the fields $F$, $\nabla\Phi$ are given by
(\ref{curv})--(\ref{covd}) respectively. 

This action is ``topological"
in the sense that it is diffeomorphism invariant; this point is
actually somewhat subtle and we return to it below. The first term of (\ref{CSredact}) is also manifestly invariant under the gauge transformations
\eq
A^h = h^{-1}\, A\, h+h^{-1}\, \dd h \ , \qquad \Phi^h= h^{-1}\, \Phi\, h
\label{APhigaugetransf}\eqend
for $h\in\Omega^0(M,\CH)$, but the second Chern--Simons type term is
generically not: Using~\cite[eq.~(3.5)]{Chamseddine:1990gk} one finds
that the full action transforms as
\eq 
{S}_{\mathrm{CSH}}^{\left(  2n-1\right)  }\big[A^h,\Phi^h \big] = {S}_{\mathrm{CSH}}^{\left(  2n-1\right)  }[A,\Phi] -2\ii (-1)^n \, \frac{(n-1)!\, n!}{(2n-1)!}\, \kappa' \, \int_M\, \Big\langle \Lambda\, \big(h^{-1}\, \dd h\big)^{2n-1} \Big\rangle \ .
\label{CStransf}\eqend
Due
to the constraint (\ref{eq20}), the closed $(2n-1)$-form $\big\langle \Lambda\, (h^{-1}\, \dd h)^{2n-1}\big\rangle$ gives a de~Rham representative for a class in the cohomology
group $H^{2n-1}(M,\pi_{2n-1}(\CH))$. Hence the deficit term in
(\ref{CStransf}) generically vanishes if and only if the free part of
the homotopy group
$\pi_{2n-1}(\CH)$ is trivial. Otherwise, the path integral
for the quantum field theory
is well-defined provided that the functional $\exp\big(\ii {S}_{\mathrm{CSH}}^{\left(
    2n-1\right)  }\big)$ is invariant under gauge transformations; this requirement generically imposes a
further topological quantization condition on the effective coupling constant
$\kappa'$ after dimensional reduction if the group
$\pi_{2n-1}(\CH)/\, {\rm Tor}(\pi_{2n-1}(\CH))$ is non-trivial.
Then up to a gauge transformation with parameter $\lambda=\xi\, \lrcorner\, A$, the infinitesimal action of diffeomorphisms of $M$ can be represented as contractions
\eq 
\delta_\xi A = \xi \, \lrcorner\, F \ , \qquad \delta_\xi \Phi=\xi \, \lrcorner\, \nabla\Phi
\label{diffgauge}\eqend
along vector fields $\xi \in \Omega^0(M,TM)$.

The field equations can be obtained by varying the reduced action (\ref{CSredact}) or equivalently by dimensional reduction over the general
condition%
\begin{equation}
\delta {S_{\rm CS}^{(2n+1)}}=\kappa\, \int_{\mathcal{M}}\,\big\langle \mathcal{F}^{n}\wedge
\delta\mathcal{A}\big\rangle =0\label{CS59}%
\end{equation}
on $\mathcal{M}= M\times S^2$. One finds that the equations of motion reduce to 
\begin{align}
\Big\langle \Big(  F^{n-1}\, \big(  2\ii\Lambda-\big[  \Phi,\Phi^{\dagger
}\big]  \big)  +\left(  n-1\right) \, F^{n-2}\wedge \nabla\Phi^{\dagger}\wedge \nabla\Phi\Big)\wedge
\delta A\Big\rangle  &  =0 \ , \nonumber \\[4pt]
\big\langle F^{n-1} \wedge \nabla\Phi^{\dagger} \ \delta\Phi\big\rangle  &  =0 \ , \nonumber \\[4pt]
\big\langle F^{n-1} \wedge \nabla\Phi \ \delta\Phi^{\dagger}\big\rangle  &  =0 \label{varphidag} \ ,
\end{align}
subject to the linear constraints (\ref{eq25})--(\ref{eq27}).
In the following we will study various aspects of the moduli space
$\CCM_n$ of
solutions to these equations modulo gauge transformations and
diffeomorphisms. As a special class of topological solutions, note
that the Higgs fields $\Phi$ are (locally) parallel
sections of the adjoint bundle ${\rm ad}(P_M)$ if and only if
the curvature two-form $F$ of $P_M$ vanishes, in which case the field
equations are immediately satisfied when $n>1$. Since in this case the diffeomorphisms
(\ref{diffgauge}) vanish on-shell, this subspace of the solution space
is the finite-dimensional moduli space of flat $\CH$-connections on $M$ modulo gauge transformations, or equivalently the moduli space of representations of the fundamental group $\pi_1(M)$ in $\CH$ modulo conjugation.

\subsection{Moduli spaces of solutions}

For some explicit examples, let us look at the case where $\CG$ is one of the classical gauge groups from section \ref{sector3}, focusing without loss of generality on $\CG=U(n)$. 
The dynamics of the reduced topological quiver gauge theory is then controlled by the invariant tensor associated to the residual gauge group (\ref{u24}).
In general, if $\{\mathsf{t}_a\}^{\dim \mathfrak{h}}_{a=1}$ denotes the generators of the Lie algebra $\mathfrak{h}$ of $\mathcal{H}$, then the invariant tensor $g_{a_{1} \cdots a_{n+1}}$ is a symmetric tensor of rank $n+1$ that is invariant under the adjoint action of $\mathcal{H}$ which we take to be the symmetrized trace~\cite{deAzcarraga:1997ya}
\begin{equation}
g_{a_{1} \cdots a_{n+1}}=\left\langle \mathsf{t}_{a_{1}} \cdots
  \mathsf{t}_{a_{n+1}} \right\rangle = \frac{1}{(n+1)!} \
\sum\limits_{\sigma\in S_{n+1}}\, \mathsf{Tr}\big(  \mathsf{t}_{a_{\sigma\left(
1\right)}  }\cdots\mathsf{t}_{a_{\sigma\left(  n+1\right) } }\big)
\label{symmtrace}\end{equation}
where $S_{n+1}$ is the symmetric group of degree $n+1$. In the
Cartan--Weyl basis the reduced gauge group $\mathcal{H}$ of
(\ref{u24}) is generated by $\{H_{i},
X_{e_{j}-e_{k}}\}_{i,j,k=1}^n$. Let us now examine in detail some cases in lower dimensionalities.

\subsubsection*{$\mbf{d=1}$}

The non-zero components of the invariant tensor for $d=1$ coincide
with the Killing--Cartan form
\begin{align}
\left\langle X_{e_{j}-e_{k}}\, X_{e_{l}-e_{m}}\right\rangle  & =%
\delta_{jm}\, \delta_{kl} \ , \nonumber \\[4pt]
\left\langle H_{i}\, X_{e_{j}-e_{k}}\right\rangle   &=\delta
_{ik}\, \delta_{ij} \ , \nonumber \\[4pt]
\left\langle H_{i}\, H_{j}\right\rangle   &=\delta_{ij} \ ,
\end{align}
and the resulting action functional is that of a topological matrix quantum mechanics given by 
\begin{equation}
S^{(1)}_{\mathrm{CSH}}=8\pi\, R^{2}\, \kappa\, \int\, \dd\tau \ \sum\limits_{\ell=0}^{m}\,
\mathsf{Tr}\big(  \phi_{(\ell+1)}\, \nabla_\tau\phi_{(\ell+1)}^{\dagger}-\phi
_{(\ell)}^{\dagger}\, \nabla_\tau\phi_{(\ell)}-2\alpha_\ell\, A_{(\ell)}\big)  \label{trivact}
\end{equation}
where $\nabla_\tau\phi_{(\ell)}=\dot{\phi}{}_{(\ell)} +A_{(\ell-1)}\,
\phi_{(\ell)}-\phi_{(\ell)}\, A_{(\ell)}$.
In this case the gauge potentials $A_{(\ell)}(\tau)\in\frh$ are Lagrange
multipliers and integrating them out of the action (\ref{trivact}) yields the constraints
\eq 
\mu_{V}^{(\ell)}(\Phi):= \phi_{(\ell+1)}\, \phi_{(\ell+1)}^{\dagger}-\phi
_{(\ell)}^{\dagger}\, \phi_{(\ell)} = 2\alpha_\ell \, \unit_{k_\ell} \ ,
\label{momentmap}\eqend
while the remaining equations of motion for the Higgs fields read $\dot{\phi}{}_{(\ell)}=0= \dot{\phi}{}_{(\ell)}^\dag$ for $\ell=0,1,\dots,m$. 

Thus in this case moduli space $\CCM_1$ of classical solutions is finite-dimensional and can be described as the subvariety cut out by the quadric (\ref{momentmap}) in the quotient of the affine variety $\prod_{\ell=0}^{m}\, \Hom(\C^{k_{\ell+1}}, \C^{k_\ell})$ by the natural action of the gauge group (\ref{u24}) given by $\phi_{(\ell+1)}\mapsto g_{\ell+1}\, \phi_{(\ell+1)}\, g_{\ell}^\dag$ with $g_\ell\in U(k_\ell)$. The moduli space $\CCM_1$ also has a representation theoretic description as an affine quiver variety in the following way. The vector space of linear representations of the $A_m$ quiver (\ref{quiver}) with fixed $V|_\CH=\bigoplus_{\ell=0}^m\, V_{k_\ell}$ is
\beq
\CCR_m(V) = \bigoplus_{\ell=0}^m\, \Hom(V_{k_{\ell+1}}, V_{k_\ell}) \ .
\eeq
The corresponding representation space for the opposite quiver, obtained by reversing the directions of all arrows, is the dual vector space $\CCR_m(V)^*$ and the cotangent bundle on $\CCR_m(V)$ is
\beq
T^*\CCR_m(V)=\CCR_m(V)\oplus \CCR_m(V)^* \ .
\eeq
It carries a canonical $\CH$-invariant symplectic structure such that
the linear $\CH$-action on $T^*\CCR_m(V)$ is
Hamiltonian~\cite{Popov:2010rf} and the corresponding moment map is
given by $\mu_V=\big(\mu_V^{(\ell)}\big)_{\ell=0}^m:T^*\CCR_m(V)\to
\frh^*$. The moduli space is then the symplectic quotient
\eq
\CCM_1= \mu_V^{-1}(2\alpha_0,2\alpha_1,\ldots,2\alpha_m)\,
\big/\!\!\big/\, \CH \ .
\label{quivervariety}\eqend
This moduli space parameterizes isomorphism classes of semisimple
representations of a certain preprojective
algebra deformed by the eigenvalues $\alpha_\ell$~\cite{Popov:2010rf}.

The topological nature of the quiver gauge theory in this instance is
not surprising as the original pure three-dimensional
Chern--Simons theory with Lagrangian
\eq
L_{\rm CS}^{(3)} = \big\langle
\alg\wedge\dd\alg+\mbox{$\frac13$}\, \alg\wedge \alg\wedge\alg 
\big\rangle
\eqend
is a topological gauge theory, and hence so is its dimensional
reduction. In this setting the affine quiver variety
(\ref{quivervariety}) is described geometrically as the
finite-dimensional moduli space of flat $SU(2)$-invariant
$\CG$-connections on the three-manifold $\CM$, which can be regarded as the symplectic quotient of the space of all $SU(2)$-invariant $\CG$-connections on $\CM$ by the action of the group of gauge transformations $\Omega^0(\CM,\CH)$.

\subsubsection*{$\mbf{d=3}$}

The Chern--Simons--Higgs like system in the case $d=3$ is the three-dimensional diffeomorphism-invariant gauge theory reduced from pure $U(n)$ Chern--Simons theory in five dimensions which has Lagrangian
\eq 
L_{\rm CS}^{(5)} = \big\langle\alg\wedge(\dd\alg)^2+\mbox{$\frac32$}\, \alg^3\wedge \dd\alg+\mbox{$\frac35$}\, \alg^5\big\rangle \ .
\eqend
As a consequence, the components of the invariant tensor are inherited from the five-dimensional theory and read as
\begin{align}
\big\langle X_{e_{j}-e_{k}}\, X_{e_{j^{\prime}}-e_{k^{\prime}}}\, X_{e_{j^{\prime
\prime}}-e_{k^{\prime\prime}}}\big\rangle  & =\delta_{kj^{\prime}}\,
\delta_{jk^{\prime\prime}}\, \delta_{k^{\prime}j^{\prime\prime}}+\delta
_{kj^{\prime\prime}}\, \delta_{jk^{\prime}}\, \delta_{k^{\prime\prime}j^{\prime}} \ , \nonumber \\[4pt]
\big\langle H_{j}\, X_{e_{j^{\prime}}-e_{k^{\prime}}}\, X_{e_{j^{\prime\prime}%
}-e_{k^{\prime\prime}}}\big\rangle  & =\delta_{jj^{\prime}}\, \delta
_{jk^{\prime\prime}}\, \delta_{k^{\prime}j^{\prime\prime}}+\delta_{jj^{\prime
\prime}}\, \delta_{jk^{\prime}}\, \delta_{k^{\prime\prime}j^{\prime}} \ , \nonumber \\[4pt]
\big\langle H_{j}\, H_{j^{\prime}}\, X_{e_{j^{\prime\prime}}-e_{k^{\prime\prime}}%
}\big\rangle  & =\delta_{jj^{\prime}}\, \big(  \delta_{jk^{\prime\prime}%
}\, \delta_{j^{\prime}j^{\prime\prime}}+\delta_{jj^{\prime\prime}}\, \delta
_{k^{\prime\prime}j^{\prime}}\big) \ , \nonumber \\[4pt]
\big\langle H_{j}\, H_{j^{\prime}}\, H_{j^{\prime\prime}}\big\rangle  &
=2\, \delta_{jj^{\prime}}\, \delta_{jj^{\prime\prime}}\, \delta_{j^{\prime}%
j^{\prime\prime}} \ .
\end{align}
With this data, the reduced action becomes
\begin{align}
S^{(3)}_{\mathrm{CSH}}=12\pi \, R^{2}\, \kappa\, \int_M \ \sum_{\ell=0}^{m}\, \mathsf{Tr}\Big( \big( & \, \phi_{(\ell+1)}\, \nabla\phi_{(\ell+1)}^{\dagger}-\phi_{(\ell)}^{\dagger
}\, \nabla\phi_{(\ell)}\big)  \wedge F_{(\ell)} \nonumber \\ & \, - 2\alpha_\ell \, A_{(\ell)}\wedge \big(
\dd A_{(\ell)}+\mbox{$\frac{2}{3}$}\, A_{(\ell)}\wedge A_{(\ell)}\big)
\Big)
\end{align}
with the field equations
\begin{align}
F_{(\ell)}\, \big(  4\alpha_{\ell}+\phi_{(\ell+1)}\, \phi_{(\ell+1)}^{\dagger}-\phi_{(\ell)}^{\dagger}\,
\phi_{(\ell)}\big) -\nabla\phi_{(\ell)}^{\dagger}\wedge \nabla\phi_{(\ell)} & =0 \ , \nonumber \\[4pt]
F_{(\ell)}\wedge \nabla\phi_{(\ell)}^{\dagger}  & =0 \ , \nonumber \\[4pt]
F_{(\ell)}\wedge \nabla\phi_{(\ell+1)}  & =0 \ .
\label{d=3equations}\end{align}

Note that the pure gauge sector of this field theory is governed by the three-dimensional Chern--Simons action with gauge group $\CH$, whose classical solution space is the moduli space of flat $\CH$-connections on $M$ modulo gauge transformations. As an explicit example, consider the case $m=1$, so that the gauge group $\CG=U(2)$ is broken to $\CH=U(1)\times U(1)$, and consider $A_1$ quiver gauge field configurations with $A_{(0)}=-A_{(1)}$ which further breaks the gauge symmetry to the diagonal $U(1)$ subgroup of $\CH$. It is then easy to reduce the field equations to the flatness conditions $F_{(0)}=-F_{(1)}=0$, and as a consequence there exists a local basis of parallel sections of the adjoint bundle ${\rm ad}(P_M)$. Hence in this case the solution space is again the finite-dimensional moduli space of flat $\CH$-connections on $M$. Owing to the topological nature of the system in this dimensionality, we believe that this is the generic moduli space of solutions in this dimension, but we have no rigorous proof of this fact.

Reduced field equations similar to those of the $m=1$ case above were obtained in~\cite{TempleRaston:1994sk}. We note that one can consider regions of $M$ with monopole type Higgs field configurations having $\nabla\Phi=0$ but $F\neq0$; in this case the monopole charge is non-zero only through two-cycles of $M$ which enclose regions where $\nabla\Phi\neq0$. According to the field equations (\ref{d=3equations}), in such regions the Higgs fields must in addition satisfy $\big[  \Phi,\Phi^{\dagger
}\big]= 2\ii\Lambda$, which is the minimum of the Higgs potential in (\ref{ymredact}). Thus monopole configurations are allowed in the Higgs vacuum and are triggered by spontaneous symmetry breaking. It would be interesting to examine the dynamics after symmetry breaking of the coupled Yang--Mills--Chern--Simons--Higgs models defined by the sum of the action functionals (\ref{ymredact}) and (\ref{CSredact}), along the lines of~\cite{Dolan:2009ie}; in this model the gauge sector also contains massive spin one degrees of freedom~\cite{Deser:1981wh}.

\subsubsection*{$\mbf{d\geq5}$}

Although for $d=3$ the moduli space of solutions is completely classified by the topology of the manifold $M$ and hence has no local degrees of freedom, in dimensions $d\geq5$ one can argue following~\cite{Banados:1995mq,Banados:1996yj,Miskovic:2005di} that the space of solutions of the diffeomorphism invariant Chern--Simons--Higgs model cannot be uniquely associated to the topology of $M$ as it generically contains local propagating degrees of freedom, depending on the algebraic properties of the invariant tensor. Our model presents an example of an \emph{irregular} Hamiltonian system~\cite{Saavedra:2000wk,Miskovic:2003ex} whose phase space is stratified into branches with different numbers of degrees of freedom and gauge symmetries, due to the dependence of the symplectic form on the fields. When certain \textit{generic} conditions are fulfilled, the symplectic form is of maximal rank and it is shown by~\cite{Banados:1996yj} using the standard Hamiltonian formalism that the number of local degrees of freedom in the pure gauge sector is given by
\begin{equation}
\mathcal{N}=\mbox{$\frac{1}{2}$} \, \big( 2(d-1)\, h-2(h+d-1)-(d-1)\, (h-1) \big) = \mbox{$\frac{1}{2}$}\, (d-1)\, (h-1) -h \ , \label{dofn}
\end{equation}
where $h>1$ is the dimension of the residual gauge group $\CH$; the first term in (\ref{dofn}) is the number of canonical variables, the second term is twice the number $h$ of first class constraints associated with the gauge symmetry plus $d-1$ first class constraints associated to spatial diffeomorphism invariance, and the third term corresponds to the second class constraints. Note that this number is zero only for $d=5$ and $h=2$, i.e. the $A_1$ quiver gauge theory in five dimensions with gauge group $\CH=U(1)\times U(1)$. 

There are also \emph{degenerate} sectors where the rank of the symplectic form is smaller, additional local symmetries emerge, and fewer degrees of freedom propagate; on these branches the constraints are functionally dependent and the standard Dirac analysis is not applicable. Thus the dynamical structure of the theory changes throughout the phase space, from purely topological sectors to sectors with the maximal number (\ref{dofn}) of local degrees of freedom. Moreover, the sector with maximal rank is stable under perturbations of the initial conditions, and on open neighbourhoods of the maximal rank solutions one can ignore the field-dependent nature of the constraints; on the contrary, degenerate sectors form measure zero subspaces of the phase space and around such degenerate backgrounds local degrees of freedom can propagate.

We do not think that this feature will be spoilt by the coupling to the Higgs fields, as the essential features should remain: The equations of motion do not constrain the connection to be flat. As our choice of invariant tensor (\ref{symmtrace}) for $\CG$ is primitive~\cite{deAzcarraga:1997ya}, we expect the generic condition to hold; note that this choice is the one that leads to the appropriate Higgs branching structure of the quiver gauge theory from section~\ref{sector3}. In fact, the phase $F=0$ is degenerate because small perturbations around it are trivial. It would be interesting to see how the degree of freedom count (\ref{dofn}) is modified by performing the analogous canonical analysis for the full Chern--Simons--Higgs model, but this seems far more complicated than the analysis of the pure Chern--Simons gauge theory. Moreover, even in the pure gauge sector, no explicit propagating solutions have been found thus far. If we choose to discard solutions with $F^n=0$, $n>1$ as degenerate backgrounds, then one can find a phase with $F$ of maximal rank which carries the maximum number of degrees of freedom (\ref{dofn}). Such a propagating phase contains ``Higgs monopole" type configurations analogous to those discussed above for the case $d=3$.

\newsection{Quiver gauge theory of AdS gravity} \label{sect4}

\subsection{$SU(2,2|1)$ Chern--Simons supergravity}

The most general action for gravity in arbitrary dimensionality is given by the
dimensional continuation of the Einstein--Hilbert action, called the
Lovelock series \cite{Lovelock:1971yv,Lanczos:1938sf,Zegers:2005vx}. In this expansion there are free
parameters which cannot be fixed from first principles. However, in
$D=2n+1$ dimensions a
special choice for the coefficients can be made in such a way that the Lovelock
Lagrangian becomes a Chern--Simons form~\cite{Chamseddine:1989nu,Zanelli:2005sa,Allemandi:2003rs,Troncoso:1999pk}. The importance of
this feature lies in the fact that the gravity theory then
possesses a gauge symmetry once the spin
connection $\omega$ and the vielbein $e$ are arranged into a
connection $\mathcal{A}$ valued in the Lie
algebra of one of the Lie groups $SO(
D-1,2)$, $SO(D,1)$ or $ISO(D-1,1)$ corresponding respectively to the local
isometry groups of
spacetimes with negative, positive or vanishing cosmological constant. Another important reason for
considering Chern--Simons gravity theories is that
they admit natural supersymmetric extensions
\cite{Chamseddine:1990gk,Troncoso:1998ng,Banados:1996hi}. In this
section we study as an example the $SU(2)$-equivariant dimensional reduction of 
five-dimensional Chern--Simons supergravity on $\CM= M\times S^2$, where
$M$ is a three-manifold.

Five-dimensional supergravity can be constructed as a Chern--Simons
gauge theory which is invariant
under the supergroup $SU(2,2|N)$~\cite{Troncoso:1997me}. The superalgebra $\mathfrak{su}%
(2,2|N)$ is the minimal supersymmetric extension of
$\mathfrak{su}(2,2)$, which is isomorphic to the
anti-de~Sitter (AdS) algebra $\mathfrak{so}(4,2)$.
A crucial observation is that in any dimension $D$ an explicit representation
of the AdS algebra can be given in terms of gamma-matrices $\Gamma_a$ which
satisfy the Clifford algebra relations (see appendix~\ref{sugrCS5})
\begin{equation}
\left\{  \Gamma_{a},\Gamma_{b}\right\}  =2\eta
_{ab}\label{cliffal}%
\end{equation}
where $\eta={\rm diag}\left(-1,1,\ldots,1\right)  $ is the metric of $D$-dimensional
Minkowski space. By defining%
\begin{equation}
\Gamma_{ab}=\mbox{$\frac{1}{2}$}\, \left[  \Gamma_{a},\Gamma
_{b}\right]
\label{Gammaab}\end{equation}
it is easy to show that%
\begin{align}
\left[  \Gamma_{a},\Gamma_{b}\right]  & =2\Gamma
_{ab}  \ , \label{gam1}\\[4pt]
\left[  \Gamma_{ab},\Gamma_{cd}\right]  & =2\left(
\eta_{cb}\, \Gamma_{ab}-\eta_{ca}\, \Gamma_{bd}+\eta
_{db}\, \Gamma_{ca}-\eta_{da}\, \Gamma_{cb}\right) \ ,  \\[4pt]
\left[  \Gamma_{ab},\Gamma_{c}\right]  & =2\left(
\eta_{cb}\, \Gamma_{a}-\eta_{ca}\, \Gamma_{b}\right) \ . \label{gam2}
\end{align}
In this way, by choosing a set of $4 \times 4$ matrices satisfying
(\ref{gam1})--(\ref{gam2}) it is possible to represent the Lie algebra
$\mathfrak{su}(2,2)  $ as a matrix algebra
by defining
\begin{equation}
\mathsf{J}_{ab}  =\mbox{$\frac{1}{2}$}\, \Gamma_{ab} \ , \qquad \mathsf{P}_{a}
=\mbox{$\frac{1}{2}$}\, \Gamma_{a} \ .
\end{equation}

Let us now turn to the supersymmetric extension
$\mathfrak{su}(2,2|N)$. For definiteness, we consider the case $N=1$
which accommodates
the minimum number $\mathcal{N}=2$ of supersymmetries. A representation of
$\mathfrak{su}(2,2|1)  $ can be obtained by extending the
bosonic generators $\left\{  \mathsf{P}_a,\mathsf{J}_{ab}\right\}  $ as
\begin{equation}
\mathsf{P}_{a}   =
\begin{pmatrix}
\frac{1}{2}\, \left(  \Gamma_{a}\right)  _{~\beta}^{\alpha} & 0\\
0 & 0
\end{pmatrix} \ , \qquad \mathsf{J}_{ab}   =
\begin{pmatrix}
\frac{1}{2}\, \left(  \Gamma_{ab}\right)  _{~\beta}^{\alpha} & 0\\
0 & 0
\end{pmatrix}
\end{equation}
and inserting the fermionic generators
\begin{equation}
\mathsf{Q}^{\gamma}  =
\begin{pmatrix}
0 & 0\\
-2\delta_{\beta}^{\gamma} & 0
\end{pmatrix} 
\ , \qquad \bar{\mathsf{Q}}_{\gamma}   =%
\begin{pmatrix}
0 & -2\delta_{\gamma}^{\alpha}\\
0 & 0
\end{pmatrix} \  .
\end{equation}
The supersymmetry algebra further requires the inclusion of a $U(1) $ generator
\begin{equation}
\mathsf{K}=%
\begin{pmatrix}
\frac{\ii}{4}\, \delta_{~\beta}^{\alpha} & 0\\
0 & \ii
\end{pmatrix}
\end{equation}
so that gauge invariance is preserved~\cite{Banados:1999kv}.

\subsection{Dimensional reduction}

In order to perform the $SU(2)$-equivariant dimensional reduction of $SU(2,2|1)  $
Chern--Simons supergravity, we choose the element
$\Lambda$ to take values in the Lorentz subalgebra $\mathfrak{so}%
(1,4)$ generated by $\left\{  \mathsf{J}_{ab}\right\}  $ and expand it as
\begin{equation}
\Lambda=\mbox{$\frac{\ii}{2}$}\, \lambda^{ab}\, \mathsf{J}_{ab} \ .
\end{equation}
This choice is not arbitrary, in the sense that it is the only one that
leads to an Einstein--Hilbert term after dimensional
reduction. Furthermore, non-trivial solutions of the constraint
equations (\ref{eq26})--(\ref{eq27}) are possible only if the Higgs
fields $\Phi$ take values in the fermionic sector of $\mathfrak{su}(
2,2|1)  $; we expand them as
\begin{equation}
\Phi  =\bar{\mathsf{Q}}_{\beta}\, \chi^{\beta} \ , \qquad \bar{\Phi}
=\bar{\chi}_{\beta}\, \mathsf{Q}^{\beta}
\end{equation}
where $\chi$ and $\bar{\chi}$ are four-component Dirac spinor zero-forms with
$\beta$ running over $1,2,3,4$. In this way the constraints
(\ref{eq26})--(\ref{eq27}) read as%
\begin{equation}
\big(  \mbox{$\frac{1}{4}$}\, \lambda^{ab}\, \left(  \Gamma_{ab}\right)  _{~\beta}^{\alpha
}+\delta_{~\beta}^{\alpha}\big) \, \chi^{\beta}=0 \ , \qquad
\bar{\chi}_{\alpha}\, \big(  \mbox{$\frac{1}{4}$}\, \lambda^{ab}\, \left(  \Gamma_{ab}\right)
_{~\beta}^{\alpha}+\delta_{~\beta}^{\alpha}\big)  =0 \ . \label{hfer}
\end{equation}
Gauging the Lie superalgebra $\mathfrak{su}(  2,2|1)  $ means that the
gauge potential decomposes as
\begin{equation}
A=\mbox{$\frac{1}{2}$}\, \omega^{ab}\, \mathsf{J}_{ab}+e^{a}\, \mathsf{P}%
_{a}+b\, \mathsf{K}+\bar{\psi}_{\alpha}\, \mathsf{Q}^{\alpha}-\bar{\mathsf{Q}%
}_{\beta}\, \psi^{\beta}%
\end{equation}
where $e, \omega$ are the standard vielbein and spin connection, $b$
is a $U(1)$ gauge field and $\psi, \bar{\psi}$ are four-component spin
$\frac32$ gravitino fields.
The constraint equation (\ref{eq25}) reads
\begin{align}
\lambda_{~b}^{a}\, \omega^{bd}  & =0 \ , \label{acons1}\\[4pt]
\lambda_{~b}^{a}\, e^{b}  & =0 \ , \\[4pt]
\bar{\psi}_{\alpha}\, \lambda^{ab}\, \left(  \Gamma_{ab}\right)  _{~\beta}^{\alpha}
& =0 \ , \qquad \\[4pt]
\lambda^{ab}\, \left(  \Gamma_{ab}\right)  _{~\beta}^{\alpha}\,
\psi^{\beta}  & =0 \ . \label{acons4}
\end{align}
These equations are still generic and will characterize the symmetry
breaking pattern once the non-zero components of $\lambda^{ab}$ are
specified. For this, we choose a particular representation of
$\mathfrak{su}(  2,2|1)  $. Using the Pauli matrices (\ref{eq17}), a representation of the Clifford algebra in five dimensions is given by 
\begin{align}
\Gamma_{0} &  =\ii\sigma_{1}\otimes\mathbbm{1}_2 \ , \label{repexpl}\\[4pt]
\Gamma_{1} &  =\sigma_{2}\otimes\mathbbm{1}_2 \ , \\[4pt]
\Gamma_{2} &  =\sigma_{3}\otimes\sigma_{1} \ , \\[4pt]
\Gamma_{3} &  =\sigma_{3}\otimes\sigma_{2} \ , \\[4pt]
\Gamma_{4} &  =\sigma_{3}\otimes\sigma_{3} \ .\label{repexpl1}
\end{align}
The explicit construction is detailed in appendix~\ref{explrep}.
We now restrict $\lambda^{ab}\, \mathsf{J}_{ab}$ to be $\lambda
^{01}\, \mathsf{J}_{01}$; other restrictions are possible and they all
lead to the same qualitative results below. With this choice the algebraic quantization condition
(\ref{eq20}) is satisfied and the constraint equation (\ref{hfer}) has non-trivial solutions if $\lambda^{01}=4$. In that case, one finds
\begin{equation}
\chi^{2}   =\chi^{4}=0=\bar{\chi}^{2}   =\bar{\chi}^{4} \ .
\end{equation}
Similarly, non-trivial solutions of (\ref{acons1})--(\ref{acons4}) are
given by taking
\begin{equation}
\omega^{1a}   =0=\omega^{0a} \ , \qquad e^{1}  =0 =e^{0} \ , \qquad
\bar{\psi}_{\alpha}  =0 =\psi^{\alpha}
\end{equation}
for $a=0,1,2,3,4$ and $\alpha=1,2,3,4$.

The reduced field content can therefore be summarised as
\begin{align}
&  e^{a},\omega^{ab} \qquad \text{for} \quad a,b=2,3,4 \ , \nonumber \\[4pt]
&  \chi_{\alpha},\bar{\chi}^{\alpha} \qquad \text{for} \quad
\alpha=1,3 \ , \nonumber \\[4pt]
&  b \qquad \text{as $U(1)$ gauge field} \ .
\end{align}
Since the reduced gauge potential becomes%
\begin{equation}
A=\mbox{$\frac{1}{2}$}\, \omega^{ab}\, \mathsf{J}_{ab}+ e^{a}\, \mathsf{P}%
_{a}+b\, \mathsf{K} \ \in \ \mathfrak{so}(  2,2)  \oplus
\mathfrak{u}(  1) \ ,
\end{equation}
the gauge symmetry $\CG=SU(2,2|1)$ is broken by this construction to
\begin{equation}
\CH=  SO(
2,2)  \times U(  1) \ .
\end{equation}
The quiver gauge
theory is thus based on the $A_1$ quiver
\begin{equation}
\xymatrix@R=0.3pc{
& \bullet \ar[r] & \bullet  }
\end{equation}
with the left node containing the $SO(2,2)$ gravitational content $e,\omega$, the
right node containing the $U(1)$ gauge field $b$, and the arrow
corresponding to the Higgs fermions $\chi$ and $\bar{\chi}$ which transform in the
bifundamental representation of $SO(2,2)\times U(1)$. Since $\pi_3(U(1))=0=\pi_3(SO(2,2))$, there is no topological quantization
condition required of the gravitational constant $\kappa'$ after
dimensional reduction.

In order to evaluate the reduced Chern--Simons--Higgs action, note
that the curvature two-form associated to the group $SO(2,2)\times U(1)$ is  
\begin{equation}
{F}  =\mbox{$\frac{1}{2}$} \, \big(
R^{ab}+\mbox{$\frac{1}{l^{2}}$} \,
e^{a}\wedge e^{b}\big) \,
\mathsf{J}_{ab}+\mbox{$\frac{1}{l}$} \, T^{a}\, \mathsf{P}_{a}+\dd b\, \mathsf{K}
\end{equation}
where $l$ is the AdS radius, $R^{ab}=\dd\omega^{ab}+\omega^a_{~c}\wedge
\omega^{cb}$ is the Lorentz curvature two-form, and $T^a=\dd
e^a+\omega^a_{~b}\wedge e^b$ is the torsion two-form.
The non-vanishing components of the $\mathfrak{su}%
(2,2|1)$-invariant tensor of rank three are given in appendix \ref{invtens}.
With this, one finds that the Chern--Simons--Higgs gravitational action is given by
\begin{equation}
{S}_{\mathrm{CSH}}^{(3)} =\frac{\kappa^\prime}{l}\, \int_{M}\, \bigg( \epsilon_{abc}\, \Big(
R^{ab}+\frac{1}{3l^{2}} \,
e^{a}\wedge e^{b}\Big)  \wedge e^{c} -\ii \nabla\bar{\chi}_{\alpha} \wedge \mathcal{Z}_{~\beta}^{\alpha
}\, \chi^{\beta}+\ii\bar{\chi}_{\alpha}\, \mathcal{Z}_{~\beta}^{\alpha} \wedge \nabla\chi^{\beta} \, \bigg) \label{CShferm}
\end{equation}
where $\kappa^\prime =8\pi \,R^{2}\, \kappa$ and
\begin{align}
\mathcal{Z}_{~\beta}^{\alpha}  & =\mbox{$\frac{1}{2}$} \, \big(
R^{ab}+\mbox{$\frac{1}{l^{2}}$} \,
e^{a}\wedge e^{b}\big) \, \left(  \Gamma_{ab}\right)
_{~\beta}^{\alpha}-\mbox{$\frac{1}{l}$}\, T^{a}\, \left(  \Gamma_{a}\right)  _{~\beta}^{\alpha}%
+\mbox{$\frac{5\ii}{2}$}\, \delta_{~\beta}^{\alpha}\, \dd b \ , \nonumber \\[4pt]
\nabla\bar{\chi}_{\alpha}  & =\dd\bar{\chi}_{\alpha}- \mbox{$\frac{1}{4}$}\,
\bar{\chi}_{\beta}\, \omega^{ab}\, \left( \Gamma_{ab}\right)  _{~\alpha}^{\beta
}-\mbox{$\frac{1}{2}$}\, \bar{\chi}_{\beta}\, e^{a}\, \left(  \Gamma_{a}\right)  _{~\alpha
}^{\beta}+\mbox{$\frac{3\ii}{4}$}\, b\, \bar{\chi}_{\beta}\, \delta_{~\alpha
}^{\beta} \ , \nonumber \\[4pt]
\nabla\chi^{\beta}  & =\dd\chi^{\beta}+\mbox{$\frac{1}{4}$}\, \omega^{ab}\, \left(  \Gamma
_{ab}\right)  _{~\alpha}^{\beta}\, \chi^{\alpha}+\mbox{$\frac{1}{2}$}\, e^{a}\, \left(
\Gamma_{a}\right)  _{~\alpha}^{\beta}\, \chi^{\alpha}-\mbox{$\frac{3\ii}{4}$}\, b\, 
\delta_{~\alpha}^{\beta} \, \chi^{\alpha} \ .
\label{eme}\end{align}
Note that the reduced field content restricts the gamma-matrices of the five-dimensional representation according to 
\begin{equation}
\Gamma_{0}=%
\begin{pmatrix}
0 & \ii\\
\ii & 0
\end{pmatrix}
\ , \qquad \Gamma_{1}=%
\begin{pmatrix}
0 & -\ii\\
\ii & 0
\end{pmatrix}
\ , \qquad \Gamma_{2}=%
\begin{pmatrix}
1 & 0\\
0 & -1
\end{pmatrix}
\ , \nonumber 
\end{equation}
\begin{equation}
\Gamma_{01}=%
\begin{pmatrix}
-1 & 0\\
0 & 1
\end{pmatrix}
\ , \qquad \Gamma_{02}=%
\begin{pmatrix}
0 & -\ii\\
\ii & 0
\end{pmatrix}
\ , \qquad \Gamma_{12}=%
\begin{pmatrix}
0 & \ii\\
\ii & 0
\end{pmatrix}
\ ,
\end{equation}
which gives a representation of the Clifford algebra in $d=2+1$
dimensions.

The infinitesimal gauge transformations corresponding to
(\ref{APhigaugetransf}) yield local symmetry transformations for
the gauge fields and Higgs fermions given by
 \begin{align}
 \delta_{\lambda,\rho}\omega^{ab}  &
 =\dd\lambda^{ab}+\omega_{~c}^{a}\, \lambda^{cb}+ \omega
 _{~c}^{b}\, \lambda^{ac}+\mbox{$\frac{1}{l^{2}}$} \, \big(
 e^{a}\wedge\rho^{b}-\rho^{a} \wedge
 e^{b}\big) \ , \\[4pt]
 \delta_{\lambda,\rho} e^{a}  & =\dd\rho^{a}+\omega_{~b}^{a}\,
 \rho^{b}-\lambda_{~b}^{a}\, e^{b} \ , \\[4pt]
 \delta_\beta b  & =\dd\beta \ , \\[4pt]
 \delta_{\rho,\kappa,\beta}\chi & =\mbox{$\frac{1}{2l}$} \, \rho^{a}\,
 \Gamma_{a}\chi- \mbox{$\frac{1}{2}$} \, \epsilon_{abc}\,
 \kappa^{ab}\, \Gamma^{c}\chi-\mbox{$\frac{3\ii}{4}$} \, \beta\, \chi \ , \label{hferm1}\\[4pt]
 \delta_{\rho,\kappa,\beta}\bar{\chi}& =-\mbox{$\frac{1}{2l}$} \, \bar{\chi}\,
 \rho^{a}\, \Gamma_{a}+\mbox{$\frac{1}{2}$} \,
 \epsilon_{abc}\, \bar{\chi}\, \kappa^{ab}\, \Gamma^{c}+
 \mbox{$\frac{3\ii}{4}$} \,
 \bar{\chi}\, \beta  \ . \label{hferm2}
 \end{align}
The action (\ref{CShferm}) describes a theory of Einstein--Hilbert
gravity with cosmological constant in three dimensions, plus a
non-minimal coupling between Higgs fermions and the fields associated
to the curvature of the residual gauge symmetry $SO(2,2)\times
U(1)$. This model is not supersymmetric as one sees from the gauge
transformations (\ref{hferm1})--(\ref{hferm2}). The equivariant
dimensional reduction scheme thus provides a novel and systematic way to couple
scalar fermions to gravitational theories, which
is normally cumbersome to do.

The variation of the Chern--Simons--Higgs action $(\ref{CShferm})$ leads to the field equations \begin{align}
2\epsilon_{abc}\check{R}^{ab}+\mbox{$\frac{\ii}{l}$}\, T_{c}\, \bar{\chi}\, \chi-\mbox{$\frac{1}{2}$}\, 
\dd b\, \bar{\chi}\, \Gamma_{c}\, \chi -\ii\nabla\bar\chi\, \wedge \Gamma_c\nabla\chi & =0 \ , \nonumber \\[4pt]
\ii\check{R}^{ab}\, \bar{\chi}\,\chi+\mbox{$\frac{1}{l}$}\, \epsilon^{abc}\, T_{c}+ \mbox{$\frac{1}{4}$}\, 
\dd b\, \bar{\chi}\, \Gamma^{ab}\chi -\ii  \nabla\bar\chi\,\wedge \Gamma^{ab}\nabla\chi & =0 \ , \nonumber \\[4pt]
\check{R}^{ab}\, \bar{\chi}\, \Gamma_{ab}\chi-\mbox{$\frac1l$}\, T^{a}\, \bar{\chi}\, \Gamma_{a}\chi
+\mbox{$\frac{15\ii}{2}$}\, \dd b\, \bar{\chi}\, \chi & =0 \ , \nonumber \\[4pt]
\mathcal{Z}\wedge\nabla\chi & =0 \ , \nonumber \\[4pt]
\nabla\bar{\chi}\wedge\mathcal{Z}  & =0 \ ,
\label{AdSmattereqs}\end{align}
where we have used the abbreviation 
\begin{equation}
\check{R}^{ab}:=\mbox{$\frac{1}{2}$}\, \big(  R^{ab}+\mbox{$\frac{1}{l^{2}}$}\, e^{a}\wedge
e^{b}\big) \ .
\end{equation}
These equations demonstrate an interesting coupling between curvature and the matter currents; note that at least one of the torsion field $T^a$ or the $U(1)$ field strength $\dd b$ must be non-zero to get a non-trivial matter coupling; otherwise, when $T^a=0=\dd b$ the matter fields freely decouple from gravity and the field equations reduce to those of pure AdS gravity in three dimensions.

\subsection{Applications}

We close with some brief discussion about possible generalizations and
applications of the gravity theories described in this
section. Five-dimensional supergravity serves as an interesting
testing ground for string theory; its Lagrangian can be obtained via
dimensional reduction of 11-dimensional supergravity where it inherits
the Chern--Simons term for the $U(1)$ gauge field $b$ from reduction
of the four-form term. Non-trivial stationary solutions of the
matter-coupled gravity theory (\ref{CShferm}) on $M$ can lead to
non-asymptotically flat stationary solutions of the original
five-dimensional supergravity theory on $\CM=M\times S^2$.  In particular, it would be interesting to seek BTZ-type solutions of the field equations (\ref{AdSmattereqs}). Note that
by restricting to the bosonic sector by setting all fermions to zero,
our reduction reduces ordinary five-dimensional AdS gravity to
three-dimensional AdS gravity without any matter fields; hence our
reduction scheme further provides a means for lifting purely
gravitational configurations on $M$ to solutions on $M\times
S^2$, and it would be interesting to examine this lifting in more
detail on some explicit solutions.

The extension of this analysis to supergroups $SU(2,2|N)$ with $N>1$
would lead to a quiver gauge theory based on the $A_{1}$ quiver with residual gauge group $\CH=SO(2,2)\times U(N)$, along with
additional $SU(N)$ gauge fields and gravitinos (see appendix~\ref{SUGRALag}). The extensions
to higher dimensions could presumably also lead to novel quiver
gauge theories of gravity-matter interactions. Our construction here is similar to the known method of compactifying Einstein--Maxwell theories over $S^2$ supported by magnetic monopole flux (see e.g.~\cite{Bouchareb:2013dka}); this technique can be used to export non-vacuum solutions with isometry group $SO(2,2)\times SO(3)$ to local $AdS_3\times S^2$ solutions of the five-dimensional Einstein equations, which in the minimal supergravity case are near-horizon limits of black strings.

It would be interesting to extend the present construction to a quiver gauge
theory of higher-spin gravity in three dimensions, which requires extending the three-dimensional Chern--Simons gauge theory
based  on the AdS group $SO(2,2)\simeq SL(2,\R)\times SL(2,\R)$ to
those based on non-compact real forms of $SL(n,\C)\times SL(n,\C)$ for $n>2$ (see
e.g.~\cite{Campoleoni:2010zq}). Finding the appropriate Chern--Simons
supergravity theory in higher dimensions could then be used as a novel
mechanism to couple matter fields to higher-spin gravity
theories. Moreover, by assigning different coupling constants to the
two $SL(2,\R)$ factors, one can couple fermionic matter fields to the
Chern--Simons gauge theory of gravity on three-dimensional
Riemann--Cartan spacetimes considered in e.g.~\cite{Cacciatori:2005wz}.

\subsection*{Acknowledgments}

We thank L.~Griguolo, F.~Izaurieta, O.~Miskovic, E.~Rodriguez, S.~Salgado and D.~Seminara for helpful discussions and correspondence.
This work was supported in part by the Consolidated Grant ST/J000310/1
from the UK Science and Technology Facilities Council. The work of O.V. is
supported by grants from the Comisi\'{o}n Nacional de Investigaci\'{o}n
Cient\'{\i}fica y Tecnol\'{o}gica \textrm{CONICYT} and from the Universidad de
Concepci\'{o}n, Chile.

\appendix

\newsection{Classical gauge groups} \label{app1}

In this appendix we summarize the group theory data which are used in
section~\ref{sector3} in the case when the gauge symmetry belongs to
one of the four infinite families $A_n,B_n,C_n,D_n$ of  classical Lie
groups in the Cartan classification; we consider each family in turn. Below $\{E_{i,j}\}_{i,j=1}^n$
denotes the orthonormal basis of $n\times n$ matrix units with elements
$\left( E_{i,j}\right) _{kl}=\delta_{ik}\, \delta_{jl}$, and $\left\{  e_{i}\right\}_{i=1}^n  $ is
the canonical orthonormal basis of~$%
\mathbb{R}
^{n}$.

\subsubsection*{$\mbf{\mathcal{G}=U(n)}$}

\begin{equation}%
\begin{tabular}
[c]{l|l|l}
Positive roots $\alpha>0$& $e_{i}-e_{j}$ & $1\leq i<j\leq n$\\\hline
Cartan generators & ${H}_{i}={E}_{i,i}$ & $1\leq i\leq
n$\\\hline
Root vectors & $X_{e_{i}-e_{j}}=E_{i,j}$ & $i\neq j$,
$i,j=1,\ldots,n$\\\hline
Weyl symmetry $\Wcal$ & $S_{n}$ & 
\end{tabular}
\ \ \ \ \ \label{u1}%
\end{equation}

\subsubsection*{$\mbf{\mathcal{G}=SO(2n+1)}  $}

\begin{equation}
\begin{tabular}
[c]{l|l|l}
Positive roots $\alpha>0$ & $e_{i}\pm e_{j}$ & $1\leq i<j\leq n$\\
& $e_{i}$ & $1\leq i\leq n$\\\hline
Cartan generators & $H_{i}=E_{i,i}-E_{i+n,i+n}$ &
$1\leq i\leq n$\\\hline
Root vectors & $X_{e_{i}-e_{j}}=E_{j+1,i+1}-E_{i+n+1,j+n+1}$ & $i\neq j$\\
& $X_{e_{i}+e_{j}}=E_{i+n+1,j+1}-E_{j+n+1,i+1}$ &
$i<j$\\
& $X_{e_{i}}=E_{1,i+1}-E_{i+n+1,1}$ & $1\leq i\leq
n$\\\hline
Weyl symmetry $\Wcal$ & $S_{n}\ltimes\left(
\mathbb{Z}
_{2}\right)  ^{n}$ &
\end{tabular}
\end{equation}

\subsubsection*{$\mbf{\mathcal{G}=Sp(2n) } $}

\begin{equation}
\begin{tabular}
[c]{l|l|l}
Positive roots $\alpha>0$ & $e_{i}\pm e_{j}$ & $1\leq i<j\leq n$\\
& $e_{2i}$ & $1\leq i\leq n$\\\hline
Cartan generators & $H_{i}=E_{i,i}-E_{i+n,i+n}$ &
$1\leq i\leq n$\\\hline
Root vectors & $X_{e_{i}-e_{j}}=E_{j,i}-E_{i+n,j+n}$ & $i\neq j$\\
& $X_{e_{i}+e_{j}}=E_{i+n,j}-E_{j+n,i}$ &
$i<j$\\
& $X_{2e_{i}}=E_{i+n,i}$ & $1\leq i\leq n$\\\hline
Weyl symmetry $\Wcal$ & $S_{n}\ltimes\left(
\mathbb{Z}
_{2}\right)  ^{n}$ &
\end{tabular}
\end{equation}

\subsubsection*{$\mbf{\mathcal{G}=SO(2n)  }$}

\begin{equation}
\begin{tabular}
[c]{l|l|l}
Positive roots $\alpha>0$ & $e_{i}\pm e_{j}$ & $1\leq i<j\leq n$\\\hline
Cartan generators & $H_{i}=E_{i,i}-E_{i+n,i+n}$ &
$1\leq i\leq n$\\\hline
Root vectors & $X_{e_{i}-e_{j}}=E_{j,i}-E_{i+n,j+n}$ & $i\neq j$\\
& $X_{e_{i}+e_{j}}=E_{i+n,j}-E_{j+n,i}$ &
$i<j$\\
& $X_{-e_{i}-e_{j}}=E_{j,i+n}-E_{i,j+n}$ &
$i<j$\\\hline
Weyl symmetry $\Wcal$ & $S_{n}\ltimes\left(
\mathbb{Z}
_{2}\right)  ^{n-1}$ &
\end{tabular}
\end{equation}

\newsection{Extended Cartan homotopy formula} \label{echfa}

\subsection{Restricted homotopy formula}

Fix $r\geq0$ and consider a set of gauge connection one-forms $\mathcal{A}_{i}$ with
$i=0,1, \dots ,r+1$ on a $D$-dimensional manifold $\mathcal{M}$,
together with a Euclidean $\left(  r+1\right)
$-simplex $\Delta_{r+1}\subset\R^{r+2}$ defined by
\begin{equation}
\Delta_{r+1}:= \Big\{ t=(t_i)_{i=0}^{r+1} \ \Big| \ t_{i} \geq0 \ , \
\mbox{$\sum\limits_{i=0}^{r+1}\, t_{i}   =1$} \Big\} \ .
\label{ch2.lalal}\end{equation}
We will sometimes write this as $\Delta_{r+1}=\langle
t_0,t_1,\ldots,t_{r+1}\rangle$. For $t\in\Delta_{r+1}$ the linear combination%
\begin{equation}
\mathcal{A}_{t}=\sum\limits_{i=0}^{r+1}\, t_{i}\, \mathcal{A}_{i}%
\end{equation}
transforms as a gauge connection in the same way as any individual form
$\mathcal{A}_{i}$. Its curvature two-form is $\mathcal{F}_t=\dd\alg_t+\alg_t\wedge\alg_t$.
Then the extended Cartan homotopy formula is given by%
\begin{equation}
\int_{\partial \Delta_{r+1}}\ \frac{\hh_{t}^{p}}{p!}
\Pi=\int_{\Delta_{r+1}}\ \frac{\hh_{t}%
^{p+1}}{\left(  p+1\right)  !} \, \dd\Pi+\left(  -1\right)  ^{p+q}\, \dd\int_{\Delta_{r+1}%
}\ \frac{\hh_{t}^{p+1}}{\left(  p+1\right)  !}\Pi \ ,
\end{equation}
where generally $\Pi$ is a polynomial in the fields $\left\{  \mathcal{A}_{t}%
,\mathcal{F}_{t},\dd_{t}\mathcal{A}_{t},\dd_{t}\mathcal{F}_{t}\right\}
$ which is simultaneously
an $m$-form on $\mathcal{M}$ and a $q$-form on $\Delta_{r+1}$ with $m\geq p$ and
$p+q=r$; here we denote by $\dd_t$ the exterior derivative on
$\Delta_{r+1}$. The operator $\hh_{t}$ is the \textit{homotopy
  derivation} which maps
differential forms on $\CM\times\Delta_{r+1}$ according to%
\begin{equation}
\hh_{t} \,:\, \Omega^{a}(  \CM)  \otimes\Omega^{b}(  \Delta_{r+1})
 \ \longrightarrow \ \Omega^{a-1}(  \CM)  \otimes\Omega^{b+1}(
\Delta_{r+1})
\end{equation}
and satisfies the Leibniz rule. The action of $\hh_{t}$ on
$\mathcal{A}_{t},\mathcal{F}_{t}$ is given by
\begin{equation}
\hh_{t}\mathcal{F}_{t}   =\dd_{t}\mathcal{A}_{t} \ , \qquad
\hh_{t}\mathcal{A}_{t}  =0 \ .
\end{equation}
The operators $\hh_{t}$, $\dd_{t}$ and $\dd$ define a graded algebra
\begin{equation}
\dd^{2}   =0=\dd_{t}^{2} \ , \qquad \left[  \hh_{t},\dd\right]
=\dd_{t} \ , \qquad \left\{  \dd,\dd_{t}\right\}     =0 = [\hh_t,\dd_t] \ .
\end{equation}

Let us look now at the particular choice of polynomial
\begin{equation}
\Pi=\left\langle \mathcal{F}_{t}^{n+1}\right\rangle \ .
\end{equation}
For this choice $\dd\Pi=0$, $q=0$ and $m=2n+2$, so that the allowed
values for $p$ are $p=0,1, \ldots,2n+2$. In this case the homotopy
formula reduces to%
\begin{equation}
\int_{\partial \Delta_{p+1}}\ \frac{\hh_{t}^{p}}{p!}\left\langle \mathcal{F}_{t}%
^{n+1}\right\rangle =\left(  -1\right)  ^{p}\, \dd\int_{\Delta_{p+1}}\
\frac{\hh_{t}^{p+1}%
}{\left(  p+1\right)  !}\left\langle
  \mathcal{F}_{t}^{n+1}\right\rangle \ ,
\label{closedhomotopy}\end{equation}
which is known as the restricted or closed version of the extended
Cartan homotopy formula.

\subsection{Chern--Weil theorem}

A well-known particular case of the homotopy formula is the Chern--Weil theorem.
Setting $p=0$ in (\ref{closedhomotopy}) gives
\begin{equation}
\int_{\partial \Delta_{1}}\, \left\langle \mathcal{F}_{t}^{n+1}\right\rangle
=\dd\int_{\Delta_{1}}\, \hh_{t}\left\langle \mathcal{F}_{t}^{n+1}\right\rangle
\label{ch2.jojo}%
\end{equation}
where $\mathcal{F}_{t}$ is the curvature of the connection $\mathcal{A}%
_{t}=t_{0}\, \mathcal{A}_{0}+t_{1}\, \mathcal{A}_{1}$ with $t_{0}+t_{1}=1$. The
boundary of the simplex $\Delta_{1}=\langle t_0,t_1\rangle
$ is given by $\partial\langle t_0,t_1\rangle=\langle
t_1\rangle-\langle t_0\rangle$ and
so the left-hand side of (\ref{ch2.jojo}) becomes
\begin{equation}
\int_{\partial \Delta_{1}}\, \left\langle \mathcal{F}_{t}^{n+1}\right\rangle
=\left\langle \mathcal{F}_{1}^{n+1}\right\rangle -\left\langle \mathcal{F}%
_{0}^{n+1}\right\rangle \ .
\end{equation}
Since $\left\langle \mathcal{F}_{t}^{n+1}\right\rangle $ is a symmetric
polynomial we have
\begin{equation}
\hh_{t}\left\langle \mathcal{F}_{t}^{n+1}\right\rangle =\left(
  n+1\right)\, 
\left\langle \hh_{t}\mathcal{F}_{t} \wedge
  \mathcal{F}_{t}^{n}\right\rangle \ ,
\end{equation}
and using
\begin{equation}
\hh_{t}\mathcal{F}_{t}  =\dd_{t}\mathcal{A}_{t}
 =\dd t_{0}\, \mathcal{A}_{0}+\dd t_{1}\, \mathcal{A}_{1}
 =\dd t_{1}\, \left(  \mathcal{A}_{1}-\mathcal{A}_{0}\right)
\end{equation}
we get
\begin{equation}
\hh_{t}\left\langle \mathcal{F}_{t}^{n+1}\right\rangle =\left(
  n+1\right) \,
\left\langle \dd t_{1}\, \big(  \mathcal{A}_{1}-\mathcal{A}_{0}\big)
\wedge \mathcal{F}_{t}^{n}\right\rangle \ .
\end{equation}
Substituting into (\ref{ch2.jojo}) one arrives at the
Chern--Weil theorem%
\begin{equation}
\left\langle \mathcal{F}_{1}^{n+1}\right\rangle -\left\langle \mathcal{F}%
_{0}^{n+1}\right\rangle =\left(  n+1\right) \, \dd\int_{0}^{1}\,
\dd t \ \big\langle
\left(  \mathcal{A}_{1}-\mathcal{A}_{0}\right) \wedge  \mathcal{F}_{t}^{n}%
\big\rangle  =\dd Q_{\mathcal{A}_{1}\leftarrow\mathcal{A}_{0}}^{\left(
    2n+1\right)  } \ .
\end{equation}

\subsection{Triangle equation} \label{triangleq}

The case $p=1$ yields the triangle equation (\ref{trieq}). Setting
$p=1$ in (\ref{closedhomotopy}) gives
\begin{equation}
\int_{\partial \Delta_{2}}\, \hh_{t}\left\langle \mathcal{F}_{t}^{n+1}\right\rangle
=-\frac{1}{2}\, \dd\int_{\Delta_{2}}\, \hh_{t}^{2}\left\langle \mathcal{F}_{t}^{n+1}%
\right\rangle \label{ch2.2klkl}%
\end{equation}
where $\mathcal{A}_{t}=t_{0}\, \mathcal{A}_{0}+t_{1}\, \mathcal{A}_{1}%
+t_{2}\, \mathcal{A}_{2}$ with $t_{0}+t_{1}+t_{2}=1$. Again the boundary of the
simplex $\Delta_{2}=\langle t_0,t_1,t_2\rangle$ is given by
$\partial\langle t_0,t_1,t_2\rangle = \langle t_1,t_2\rangle -\langle
t_0,t_2\rangle +\langle t_0,t_1\rangle$ and so the left-hand side of
(\ref{ch2.2klkl}) becomes
\begin{equation}
\int_{\partial \Delta_{2}}\, \hh_{t}\left\langle \mathcal{F}_{t}^{n+1}\right\rangle
=Q_{\mathcal{A}_{2}\leftarrow\mathcal{A}_{1}}^{\left(  2n+1\right)
}-Q_{\mathcal{A}_{2}\leftarrow\mathcal{A}_{0}}^{\left(  2n+1\right)
}+Q_{\mathcal{A}_{1}\leftarrow\mathcal{A}_{0}}^{\left(  2n+1\right)
} \ , \label{ch2.2.ttt}
\end{equation}
where we have used
\begin{equation}
\int_{\langle t_i,t_j\rangle  }\, \hh_{t}\left\langle
\mathcal{F}_{t}^{n+1}\right\rangle =Q_{\mathcal{A}_{j}\leftarrow
\mathcal{A}_{i}}^{\left(  2n+1\right)  } \ .
\end{equation}
Using the symmetry of the invariant polynomial $\left\langle - \right\rangle $ one
derives%
\begin{equation}
\mbox{$\frac{1}{2}$}\, \hh_{t}^{2}\left\langle
  \mathcal{F}_{t}^{n+1}\right\rangle = \mbox{$\frac
{1}{2}$}\, n\, \left(  n+1\right) \, \big\langle \left(  \dd_{t}\mathcal{A}_{t}\right)
^{2} \wedge \mathcal{F}_{t}^{n+1}\big\rangle \label{ch2.2.popo}%
\end{equation}
where%
\begin{equation}
\dd_{t}\mathcal{A}_{t}=\dd t_{0}\, \mathcal{A}_{0}+\dd t_{1}\, \mathcal{A}_{1}%
+\dd t_{2}\, \mathcal{A}_{2}=\dd t_{0}\, \left(  \mathcal{A}_{0}-\mathcal{A}_{1}\right)
+\dd t_{2}\, \left(  \mathcal{A}_{2}-\mathcal{A}_{1}\right) \ .
\end{equation}
Substituting into (\ref{ch2.2.popo}) gives
\begin{equation}
\mbox{$\frac{1}{2}$}\, \hh_{t}^{2}\left\langle
  \mathcal{F}_{t}^{n+1}\right\rangle =-n\, \left(
n+1\right) \, \dd t_{0}\wedge \dd t_{2}\, \left\langle \left(  \mathcal{A}_{2}-\mathcal{A}%
_{1}\right) \wedge \left(  \mathcal{A}_{1}-\mathcal{A}_{0}\right)  \wedge \mathcal{F}%
_{t}^{n+1}\right\rangle \ .
\end{equation}
We redefine the simplex parameterization as $t   =1-t_{0}$, $s  =t_{2}$
and integrate explicitly over $\Delta_{2}$. In this way we get
\begin{equation}
\frac{1}{2}\, \int_{\partial \Delta_{2}}\, \hh_{t}^{2}\left\langle \mathcal{F}_{t}%
^{n+1}\right\rangle =Q_{\mathcal{A}_{2}\leftarrow\mathcal{A}_{1}%
\leftarrow\mathcal{A}_{0}}^{\left(  2n\right)  }%
\end{equation}
where $Q_{\mathcal{A}_{2}\leftarrow\mathcal{A}_{1}\leftarrow\mathcal{A}_{0}%
}^{\left(  2n\right)  }$ is defined in (\ref{Q2n}). Substituting this
expression together with (\ref{ch2.2.ttt}) into (\ref{ch2.2klkl}) we
arrive finally at the triangle equation (\ref{trieq}).

\newsection{$ {SU(  2,2|N) } $
Chern--Simons supergravity}\label{sugrCS5}

\subsection{Supergravity Lagrangian}\label{SUGRALag}

The supersymmetric extension of the AdS algebra in five dimensions is the Lie superalgebra $\mathfrak{su}(  2,2|N)  $~\cite{Chamseddine:1990gk}. The
associated gauge field decomposes into generators as
\begin{equation}
A=e^{a}\, \mathsf{P}_{a}+\mbox{$\frac{1}{2}$}\, \omega^{ab}\, \mathsf{J}_{ab}%
+a_{~n}^{m}\, \mathsf{M}_{~m}^{n}+b\, \mathsf{K}+\bar{\psi}_{\alpha}^{k}%
\, \mathsf{Q}_{k}^{\alpha}-\bar{\mathsf{Q}}_{\beta}^{k}\, \psi_{k}^{\beta} \ .
\end{equation}
Here the generators $\left\{ \mathsf{P}_a, \mathsf{J}_{ab} \right\}$ span
an $\mathfrak{so}(4,2)  $ subalgebra, $\mathsf{M}_{~m}^{n}$ are
$N^{2}-1$ generators of $SU(  N)  $, $\mathsf{K}$
generates a $U(1)  $ subgroup, and $\mathsf{Q}_{k}^{\alpha
},\bar{\mathsf{Q}}_{\beta}^{k}$ are the supersymmetry generators. The
Chern--Simons Lagrangian associated to this superalgebra is given by \cite{Chamseddine:1990gk,Troncoso:1998ng,Izaurieta:2006wv}
\begin{equation}
L_{\text{CS}}^{\left(  5\right)  }=L_{\psi}
+L_{a}+L_{b}+L_{e}
\end{equation}
where%
\begin{align}
L_{\psi}&  =\mbox{$\frac{3}{2\ii}$}\, \left(  \bar{\psi}^n%
\wedge \mathcal{R}\wedge \nabla\psi_n +\bar{\psi}^{n}\wedge \mathcal{F}_{~n}^{m}\wedge \nabla\psi_{m}%
-\nabla\bar{\psi}^n\wedge \mathcal{R}\wedge \psi_n -\nabla\bar{\psi}^{n}\wedge \mathcal{F}_{~n}^{m}%
\wedge \psi_{m}\right) \ , \nonumber \\[4pt]
L_{a} &  =\mbox{$\frac{3}{N}$}\, \dd b \wedge
\mathsf{Tr}\big(  a\wedge \dd a+\mbox{$\frac{2}{3}$}\, a^{3}\big)  -\ii \mathsf{Tr}\big( a\wedge \left(
\dd a\right)  ^{2}+\mbox{$\frac{3}{2}$}\, a^{3}\wedge \dd a+\mbox{$\frac{3}{5}$}\, a^{5}\big) \ , \nonumber \\[4pt]
L_{b} &  = \big( \mbox{$ \frac{1}{16}-\frac{1}
{N^{2}}$} \big) \, b\wedge \left(  \dd b\right)  ^{2}-\mbox{$\frac{3}{4l^{2}}$}\, b\wedge \big(
T^{a}\wedge T_{a}-R_{ab}\wedge e^{a}\wedge e^{b}-\mbox{$\frac{l^{2}}{2}$}\, R^{ab}\wedge R_{ab}\big) \ , \nonumber \\[4pt]
L_{e} &  =\mbox{$\frac{3}{8l}$}\, \epsilon_{abcdh}\, \big(
R^{ab}\wedge R^{cd}+\mbox{$\frac{2}{3}$}\, R^{ab}\wedge e^{c}\wedge e^{d}+\mbox{$\frac{1}{5}$}\, e^{a}\wedge e^{b}\wedge e^{c}\wedge
e^{d}\big)\wedge e^{h} \ ,
\end{align}
and%
\begin{align}
\mathcal{R} &  =\ii \big( \mbox{$ \frac{1}{4}+\frac{1}{N}$} \big)\,
\big( \dd b+\mbox{$\frac{\ii}{2l}$}\, \bar{\psi}^n\wedge \psi_n \big) + \mbox{$\frac{1}{2}$}\, \big(  T^{a}%
-\mbox{$\frac{1}{4}$}\, \bar{\psi}^n\wedge \Gamma^{a}\psi_n \big)\, \Gamma_{a} \nonumber \\ & \qquad +\, \mbox{$\frac{1}{4}$}\, \big(
R^{ab}+\mbox{$\frac{1}{l}$}\, e^{a}\wedge
e^{b}+\mbox{$\frac{1}{4l}$}\, \bar{\psi}^n\wedge \Gamma^{ab}\psi_n \big)\,
\Gamma_{ab} \ , \nonumber \\[4pt]
\mathcal{F}_{~n}^{m} & =f_{~n}^{m}-\mbox{$\frac{1}{2l}$}\, \bar{\psi}^{m}\wedge \psi_{n} \ .
\end{align}
Here the spinor covariant derivatives are defined by
\begin{align}
\nabla\psi_k& =\dd \psi_k+\mbox{$\frac1{2l}$}\, e^a\wedge \Gamma_a\psi_k+\mbox{$\frac14$} \, \omega^{ab}\wedge\Gamma_{ab}\psi_k -a^n_{~k}\wedge \psi_n+\ii\big(\mbox{$\frac14-\frac1N$} \big)\, b\wedge\psi_k \ , \nonumber \\[4pt]
\nabla\bar{\psi}^k& =\dd \bar{\psi}^k-\mbox{$\frac1{2l}$}\, e^a\wedge \bar{\psi}^k\Gamma_a-\mbox{$\frac14$} \, \omega^{ab}\wedge\bar{\psi}^k\Gamma_{ab}+ a^k_{~n}\wedge \bar{\psi}^n-\ii\big(\mbox{$\frac14-\frac1N$} \big)\, b\wedge\bar{\psi}^k \ ,
\end{align}
while $f=\dd a+a\wedge a$ is the curvature of the $SU(N)$ gauge field $a$. The (super)symmetry transformations and field equations can be read off from the general expressions (\ref{calAgaugetransf}) and (\ref{calFfieldeq}) respectively.

\subsection{Representation of $\mathfrak{su}(  2,2|1)  $} \label{explrep}

For simplicity we consider now the particular instance
$N=1$. This case furnishes the minimum number $\mathcal{N} =2$ of supersymmetries, and the commutation relations are given by%
\begin{align}
\left[  \mathsf{K},\mathsf{Q}^{\rho}\right]   &  =\mbox{$\frac{3\ii}{4}$}\, \mathsf{Q}%
^{\rho} \ , \nonumber \\[4pt]
\left[  \mathsf{K},\bar{\mathsf{Q}}_{\rho}\right]   &  =-\mbox{$\frac{3\ii}%
{4}$}\, \bar{\mathsf{Q}}_{\rho} \ , \nonumber \\[4pt]
\left[  \mathsf{P}_{a},\mathsf{P}_{b}\right]   &  = \mbox{$\frac{1}{l^{2}}$}\,
\mathsf{J}_{ab} \ , \nonumber \\[4pt]
\left[  \mathsf{P}_{a},\mathsf{J}_{bc}\right] &  =\eta_{ba}\, \mathsf{P}%
_{c}-\eta_{ac}\, \mathsf{P}_{b} \ , \nonumber \\[4pt]
\left[  \mathsf{P}_{a},\mathsf{Q}^{\rho}\right]   &  =-\mbox{$\frac{1}{2l}$}\, \left(
\Gamma_{a}\right)  _{~\gamma}^{\rho}\, \mathsf{Q}^{\gamma} \ , \nonumber \\[4pt]
\left[  \mathsf{P}_{a},\bar{\mathsf{Q}}_{\rho}\right]   &  =\mbox{$\frac{1}%
{2l}$}\, \bar{\mathsf{Q}}_{\gamma}\, \left(  \Gamma_{a}\right)  _{~\rho}^{\gamma} \ , \nonumber \\[4pt]
\left[\mathsf{J}_{ab},\mathsf{J}_{cd}\right] &  =\eta_{cb}\, \mathsf{J}%
_{ad}-\eta_{ac}\, \mathsf{J}_{bd}+\eta_{db}\, \mathsf{J}_{ca}-\eta_{ad}%
\, \mathsf{J}_{cb} \ , \nonumber \\[4pt]
\left[  \mathsf{J}_{ab},\mathsf{Q}^{\rho}\right]   &  =- \mbox{$\frac{1}{2}$}\, \left(
\Gamma_{ab}\right)  _{~\gamma}^{\rho}\, \mathsf{Q}^{\gamma} \ , \nonumber \\[4pt]
\left[  \mathsf{J}_{ab},\bar{\mathsf{Q}}_{\rho}\right]   &  =\mbox{$\frac{1}%
{2}$}\, \bar{\mathsf{Q}}_{\gamma}\left(  \Gamma_{ab}\right)  _{~\rho}^{\gamma} \ , \nonumber \\[4pt]
\left\{  \mathsf{Q}^{\rho},\bar{\mathsf{Q}}_{\sigma}\right\}   &
=-4\ii\delta_{~\sigma}^{\rho}\,\mathsf{K}+2\left(  \Gamma^{a}\right)  _{~\sigma
}^{\rho}\, \mathsf{P}_{a}-\left(  \Gamma_{ab}\right)  _{~\sigma}^{\rho}%
\, \mathsf{J}_{ab} \ .
\end{align}
According to (\ref{repexpl})--(\ref{repexpl1}) the matrix
generators explicitly read as
\begin{equation}
\Gamma_{0}=%
\begin{pmatrix}
0 & \ii & 0 & 0\\
\ii & 0 & 0 & 0\\
0 & 0 & 0 & -\ii\\
0 & 0 & -\ii & 0
\end{pmatrix} \ , \qquad \Gamma_{1}=%
\begin{pmatrix}
0 & -\ii & 0 & 0\\
\ii & 0 & 0 & 0\\
0 & 0 & 0 & \ii\\
0 & 0 & -\ii & 0
\end{pmatrix} \ , \qquad \Gamma_{2}=%
\begin{pmatrix}
1 & 0 & 0 & 0\\
0 & -1 & 0 & 0\\
0 & 0 & -1 & 0\\
0 & 0 & 0 & 1
\end{pmatrix} \ ,
\nonumber \end{equation}
\begin{equation}
\Gamma_{3}=%
\begin{pmatrix}
0 & 0 & 1 & 0\\
0 & 0 & 0 & 1\\
1 & 0 & 0 & 0\\
0 & 1 & 0 & 0
\end{pmatrix}
\ , \qquad \Gamma_{4}=%
\begin{pmatrix}
0 & 0 & -\ii & 0\\
0 & 0 & 0 & -\ii\\
\ii & 0 & 0 & 0\\
0 & \ii & 0 & 0
\end{pmatrix} \ ,
\end{equation}
and using (\ref{Gammaab}) we find
\begin{equation}
\Gamma_{01}=%
\begin{pmatrix}
-1 & 0 & 0 & 0\\
0 & 1 & 0 & 0\\
0 & 0 & -1 & 0\\
0 & 0 & 0 & 1
\end{pmatrix}
\ , \qquad \Gamma_{02}=%
\begin{pmatrix}
0 & -\ii & 0 & 0\\
\ii & 0 & 0 & 0\\
0 & 0 & 0 & -\ii\\
0 & 0 & \ii & 0
\end{pmatrix}
\ , \qquad \Gamma_{03}=%
\begin{pmatrix}
0 & 0 & 0 & \ii\\
0 & 0 & \ii & 0\\
0 & -\ii & 0 & 0\\
-\ii & 0 & 0 & 0
\end{pmatrix} \ ,
\nonumber \end{equation}
\begin{equation}
\Gamma_{04}=%
\begin{pmatrix}
0 & 0 & 0 & 1\\
0 & 0 & 1 & 0\\
0 & 1 & 0 & 0\\
1 & 0 & 0 & 0
\end{pmatrix}
\ , \qquad \Gamma_{12}=%
\begin{pmatrix}
0 & \ii & 0 & 0\\
\ii & 0 & 0 & 0\\
0 & 0 & 0 & \ii\\
0 & 0 & \ii & 0
\end{pmatrix}
\ , \qquad \Gamma_{13}=%
\begin{pmatrix}
0 & 0 & 0 & -\ii\\
0 & 0 & \ii & 0\\
0 & \ii & 0 & 0\\
-\ii & 0 & 0 & 0
\end{pmatrix} \ ,
\nonumber \end{equation}
\begin{equation}
\Gamma_{14}=%
\begin{pmatrix}
0 & 0 & 0 & -1\\
0 & 0 & 1 & 0\\
0 & -1 & 0 & 0\\
1 & 0 & 0 & 0
\end{pmatrix}
\ , \qquad \Gamma_{23}=%
\begin{pmatrix}
0 & 0 & 1 & 0\\
0 & 0 & 0 & -1\\
-1 & 0 & 0 & 0\\
0 & 1 & 0 & 0
\end{pmatrix}
\ , \qquad \Gamma_{24}=%
\begin{pmatrix}
0 & 0 & -\ii & 0\\
0 & 0 & 0 & \ii\\
-\ii & 0 & 0 & 0\\
0 & \ii & 0 & 0
\end{pmatrix} \ ,
\nonumber \end{equation}
\begin{equation}
\Gamma_{34}=%
\begin{pmatrix}
\ii & 0 & 0 & 0\\
0 & \ii & 0 & 0\\
0 & 0 & -\ii & 0\\
0 & 0 & 0 & -\ii
\end{pmatrix} \ .
\end{equation}
It is then easy to show that this particular choice of basis for the Lie
algebra $\mathfrak{su}(2,2)$ has traceless generators all satisfying the Clifford algebra relations (\ref{cliffal}).

The $\mathfrak{su}(
2,2| 1)  $-invariant tensor of rank three can be computed from this
representation as the supersymmetrized supertraces of products of
triples of supermatrices. The non-vanishing components are given by~\cite{Izaurieta:2006wv}
\begin{align}
\left\langle \mathsf{J}_{ab}\, \mathsf{J}_{cd}\, \mathsf{P}_{e}\right\rangle  &
=-\mbox{$\frac{\gamma}{2l}$}\, \epsilon_{abcde} \ , \nonumber \\[4pt]
\left\langle \mathsf{K\, K\, K}\right\rangle  &  =-\mbox{$\frac{15}{16}$} \ , \nonumber \\[4pt]
\left\langle \mathsf{K\, P}_{a}\, \mathsf{P}_{b}\right\rangle  &
=-\mbox{$\frac{1}{4l^{2}}$}\, \delta_{ab} \ , \nonumber \\[4pt]
\left\langle \mathsf{J}_{ab}\, \mathsf{K\, J}_{cd}\right\rangle  &
=-\mbox{$\frac{1}{4}$}\, (\delta_{ad}\, \delta_{bc}+\delta_{ac}\, \delta_{bd}) \ , \nonumber \\[4pt]
\left\langle \mathsf{Q}^{\alpha}%
\, \mathsf{K\, \bar{Q}}_{\beta}\right\rangle  &  =\mbox{$\frac{5}{2l}$}\, \delta^\alpha_{~\beta} \ , \nonumber \\[4pt]
\left\langle \mathsf{Q}^{\alpha
}\, \mathsf{P}_{a}\, \bar{\mathsf{Q}}_{\beta}\right\rangle  &  =-\mbox{$\frac{\ii}{l}$}\, 
\left(  \Gamma_{a}\right)  _{~\beta}^{\alpha}%
\ , \nonumber \\[4pt]
\left\langle \mathsf{Q}^{\alpha
}\, \mathsf{J}_{ab}\, \bar{\mathsf{Q}}_{\beta}\right\rangle  &  =-\mbox{$\frac{\ii}{l}$}\,
\left(  \Gamma_{ab}\right)  _{~\beta}^{\alpha}%
 \ ,
\end{align} \label{invtens}
where $\gamma$ is an arbitrary constant.

\bibliographystyle{utphys}

\providecommand{\href}[2]{#2}\begingroup\raggedright\endgroup

\end{document}